\documentclass[aps,superscriptaddress,showpacs,prd,nofootinbib,eqsecnum]{revtex4-1}

\usepackage{graphicx}
\graphicspath{ {plots/} }
\usepackage{latexsym} 
\usepackage{epsfig} 
\usepackage{xcolor}
\usepackage{notoccite}
\usepackage{subfigure}
\usepackage{graphicx,slashed,bbold,subfigure,amsmath,amssymb,enumerate}
\usepackage{natbib} 

\usepackage{ulem} 

\usepackage[colorlinks=true,citecolor=blue,linkcolor=blue,urlcolor=blue]{
hyperref}
\setcitestyle{square, numbers, comma, sort&compress}

\begin{document}

\title{ Regularizing  thermo and magnetic contributions  within nonrenormalizable theories  }

\author{Sidney S. Avancini}\email{sidney.avancini@ufsc.br}
\affiliation{Departamento de F\'{\i}sica, Universidade Federal de Santa
  Catarina, Florian\'{o}polis, SC 88040-900, Brazil}  
\author{ Ricardo L. S. Farias}\email{ ricardo.farias@ufsm.br}
\affiliation{Departamento de F\'{\i}sica, Universidade Federal de Santa
  Maria, Santa Maria, RS 97105-900, Brazil}
\author{Marcus B. Pinto} \email{marcus.benghi@ufsc.br}
\affiliation{Departamento de F\'{\i}sica, Universidade Federal de Santa
  Catarina, Florian\'{o}polis, SC 88040-900, Brazil}  
  
\author{Tulio E. Restrepo} \email{tulio.restrepo@posgrad.ufsc.br}
\affiliation{Departamento de F\'{\i}sica, Universidade Federal de Santa
  Catarina, Florian\'{o}polis, SC 88040-900, Brazil}
\affiliation{CFisUC - Center for Physics of the University of Coimbra, Department of Physics, Faculty of Sciences and Technology, University of Coimbra, 3004-516 Coimbra, Portugal}  
 
\author{William R. Tavares} \email{william.tavares@posgrad.ufsc.br}
\affiliation{Departamento de F\'{\i}sica, Universidade Federal de Santa
  Catarina, Florian\'{o}polis, SC 88040-900, Brazil}

 \begin{abstract} 
The importance of implementing a proper regularization procedure in order to treat  thermo and magnetic contributions within nonrenormalizable theories is investigated.  Our study suggests  that  potential divergences  should   be  isolated into the vacuum and purely magnetic contributions and  then  regularized  while the   convergent  thermomagnetic contributions should  be integrated over the full momentum range.  This prescription is illustrated by applying the proper time formalism to the two flavor Polyakov--Nambu--Jona-Lasinio model, whose magnetic field dependent coupling has been recently determined. Observables such as the pressure, magnetization,  speed of sound squared, and  specific heat  evaluated within our scheme are compared with results furnished by other three possible  prescriptions. We show that these quantities display   a thermomagnetic behavior  which is physically more consistent when our scheme is adopted. In particular, we  demonstrate that  naively regulating  the (entangled) vacuum, magnetic and thermomagnetic contributions leads to physically inconsistent results especially at the high temperature domain. 
 \end{abstract}

 
\maketitle
\section{Introduction}

Understanding the behavior of magnetized quark matter is of fundamental importance for the correct description of physical situations that may take place in magnetars as well as in peripheral heavy ion collisions \cite{Duncan:1992hi, Fukushima:2008xe,Kharzeev:2009pj,Kharzeev:2009mf}.  On the theoretical side  this problem has received a lot of attention recently through the use  lattice QCD (LQCD) numerical simulations and analytical model approximations. One controversy has emerged when precise LQCD applications, carried out at zero baryonic densities and physical pionic masses, have indicated that the crossover pseudocritical temperature ($T_{pc}$) should decrease with increasing magnetic field values \cite{Bali:2012zg,Bandyopadhyay:2020zte}. This outcome contradicts  early LQCD simulations \cite{DElia:2010abb,DElia:2011koc,Ilgenfritz:2012fw}, where high pionic mass values were considered,\footnote{ In a recent LQCD study \cite{Endrodi:2019zrl}, the authors obtain the decrease of $T_{pc}$ with  increasing magnetic fields when heavy pions are considered.}  and also  analytical evaluations (mostly at the mean field level)  performed with effective theories such as  the  Nambu--Jona-Lasinio model (NJL) \cite{Nambu:1961tp,*Nambu:1961fr} and the quark meson model (QMM)\cite{Scavenius:2000qd,Fraga:2008qn} as well as  their Polyakov loop extended versions (PNJL \cite{Ratti:2005jh} and PQMM \cite{Schaefer:2009ui} respectively). The possible  origin of this discrepancy has later been elucidated in Ref. \cite {Fukushima:2012kc} where the authors have shown that adding pionic loops to the mean field quark pressure could fix  the problem. Another possible alternative is to include thermomagnetic  effects on the coupling constants~\cite{Farias:2014eca, Ferrer:2014qka,Ayala:2018wux,Ayala:2017thy,Ayala:2016bbi,Ayala:2015bgv,Ayala:2015lta}. This approach has been adopted in various investigations where different possible {\it ansatz} \cite{Farias:2016gmy,Endrodi:2019whh,Ferreira:2014kpa,Ferreira:2013tba} to describe the $B$-dependent  running of the coupling have been proposed  leading to results for $T_{pc}$ which are in line  with  LQCD predictions. The same type of technique  has  been generalized  to the SU(3) PNJL model where the six fermion coupling has also been fixed according to  LQCD data \cite{Moreira:2020wau}. The reader is referred to Ref. \cite{Andersen:2014xxa} for a comprehensive  review on effective models under strong magnetic fields.

Recently, the two-flavor PNJL model coupling has been fixed by using, as input, LQCD results that determine the baryon spectrum of 1+1+1-flavors in the presence of a strong magnetic background and at physical pionic masses \cite {Endrodi:2019whh}. In this case, the running coupling, $G(B)$, has been determined by constraining  the PNJL constituent quark mass  to match the LQCD  results including, of course, the decrease of $T_{pc}$ with $B$. For the present work, it is important to remark that to evaluate quantities such as the pressure the authors of Ref. \cite {Endrodi:2019whh} have adopted Schwinger's proper time formalism \cite{Schwinger:1951nm} and regularized {\it all} integrals without separating the divergent (vacuum) piece from the convergent (thermomagnetic) contribution. Next, this regularization choice   has been used to evaluate the effective mass, the quark condensate, and the pressure for $eB=0$ to $0.6 \;{\rm GeV}^2$ covering temperatures up to $T \approx 0.27\; {\rm GeV}$ so that the final result for  $T_{pc}(B)$ is in good agreement with LQCD predictions. Nevertheless, it is now well established that such a regularization procedure can give rise to a series of inconsistencies if the calculations are pushed to higher temperatures (when, e.g., the pressure may not converge to the Stefan-Boltzmann limit, as observed in the case $eB = 0$ \cite{Costa:2009ae,Ruivo:2010ff,Costa:2010zw}) or to finite baryon chemical potentials (where the appearance of unphysical oscillations is more noticeable). For example, using a $B$-dependent regularization scheme for the calculation of the magnetization, some authors find oscillations  which  are  unphysical while others  find  imaginary  meson  masses which  are  in  fact  spurious  solutions  due to  the  inappropriate  choice  of  the  regularization. This can  be  seen  by  comparing  the  meson  masses  results  of Ref. \cite{Fayazbakhsh:2012vr} where    real  values can only be  reproduced  upon implementing  a  consistent  regularization  strategy.  

The importance of using an appropriate regularization scheme to describe magnetized  quark  matter  has  been  clearly  demonstrated in Refs. \cite{Avancini:2019wed,Allen:2015paa,Duarte:2015ppa} which  suggest a  strong  dependence on the choice of the regularization scheme for  the calculation of physical observables and, more importantly, that an inappropriate regularization scheme may give rise to spurious solutions. 
Here,  our main goal is to investigate different regularization schemes in order to select the ones which produce physically more reliable results as far as nonrenormalizable theories are concerned. With this aim it is important to first recall  the possibility that different  regularization (and renormalization) procedures implemented  to treat divergences in quantum field theories always introduce some degree of arbitrariness during formal evaluations. Generally, within renormalizable theories such as QCD this arbitrariness is associated to the possibility of choosing  different regulators, subtraction points, and  energy scales.  However, physically unambiguous results may be obtained by further constraints such as the ones imposed by the renormalization group equations which require that physical observables be invariant with respect to changes in the arbitrary energy scale. The situation is less clear when it comes to nonrenormalizable theories such as the PNJL model in $3+1d$ considered here. Traditionally, this type of theory is regularized with a sharp cutoff \footnote {See Ref. \cite{Inagaki:2012re} for alternatives such as dimensional regularization.}, $\Lambda$, which instead of being removed by some subtraction prescription is formally treated as a ``parameter" that sets the energy scale value (generally  $\sim 0.6-1 \, {\rm GeV}$)  up to which the theory can be considered to be effective \cite{BUBALLA_2005}.  In the absence of control parameters such as $T$, $\mu$, and $B$, where it is free from ambiguities, this became the standard procedure to deal with NJL type of models. The case of $T=B=0$ and $\mu \ne 0$ is also unambiguous since the Fermi momentum, $p_F = (\mu^2 - M^2)^{1/2}$, naturally regularizes the convergent contributions. When finite temperatures come into play (still at $B=0$), the thermodynamical potential at the (one loop) mean field level considered here splits into two parts, $\Omega = \Omega_V + \Omega_T$. The first represents the ultraviolet (UV)  divergent vacuum piece and the second one represents the convergent thermal contribution. In renormalizable theories, one usually isolates and renormalizes the divergences contained in $\Omega_V$ so that its final contribution is finite in the extreme UV limit and does not depend on any regulator. At the same time, the convergent  thermal integrals are simply integrated over the full momentum range. When considering nonrenormalizable theories where the final vacuum contribution depends on the  regulator (now a finite valued parameter), one  has two options to treat the thermal part. In early works (see Ref. \cite {Zhuang:1994dw} and references therein), the preferred method was that (for ``consistency") one should also regularize the convergent thermal integrals so that $\Omega = \Omega_V(\Lambda) + \Omega_T(\Lambda)$.  Later, it has been suggested that this course of action is unnecessary since the finite temperature contribution has a natural cutoff in itself specified by the temperature \cite{Fukushima:2003fw} and in this case one should consider $\Omega = \Omega_V(\Lambda) + \Omega_T(\infty)$. The drawback associated with the former approach is that thermodynamical quantities evaluated in this way do not converge to the expected Stefan-Boltzmann limit as $T \to \infty$ \cite {Zhuang:1994dw}. On the other hand, although the latter strategy does not spoil the high-$T$  behavior of quantities such as the pressure, it can generate thermal effective masses which are smaller than the actual current mass value, $m_c$.  Although happening at temperatures of the order $T \approx M(0) \approx 300\;{\rm MeV}$, this is still an undesirable feature \cite{Costa:2010zw}. It should be mentioned  that alternatives to recover the Stefan-Boltzmann limit while maintaining $M(T) > m_c$ at high-$T$ have been given in Refs. \cite{Bratovic:2012qs,Moreira:2010bx}.  

Additional care is needed when regularizing  a nonrenormalizable theory  to describe magnetized hadronic matter due to the  
Landau level (LL) structure acquired by the vacuum energy.  To treat this situation, a seminal work \cite {Klevansky:1989vi} employed the proper time formalism without disentangling the purely magnetic part from the vacuum so that $\Omega = \Omega_{VM}(\Lambda)$.  Adding finite temperature effects  brings in, once again, the question about the need to regularize (or not) the finite thermomagnetic contribution, $\Omega_{TM}$. Since these two possibilities will be examined here, let us dub standard proper time (SPT) the one in which this term is {\it not} regularized so that  $\Omega = \Omega_{VM}(\Lambda) + \Omega_{TM}(\infty)$. At the same time, let us dub thermomagnetic regulated  proper time scheme (TRPT), the one in which  $\Omega_{TM}$  is regularized as in Ref. \cite {Endrodi:2019whh} so that $\Omega = \Omega_{VM}(\Lambda) + \Omega_{TM}(\Lambda)$. Later, the  interesting possibility  of isolating all the  divergences within the vacuum term by separating a finite purely magnetic term (summed over all LL) has been suggested \cite {Ebert:2003yk,Ebert:1999ht}. This scheme, which avoids unphysical oscillations, was originally applied in the proper time (PT) framework \cite {Ebert:1999ht} and latter generalized to the sharp cutoff framework \cite {Menezes:2008qt} allowing for several further applications \cite{Avancini:2012ee,Menezes:2009uc,Avancini:2016fgq,Avancini:2017gck,Avancini:2018svs,Coppola:2017edn,Duarte:2015ppa,Allen:2015paa,Duarte:2016pgi}.  Within this magnetic field {\it independent} regularization scheme\footnote {More recently, the MFIR has been further improved by means of the Hurwitz-Riemann-zeta function defining the so called zeta MFIR regularization procedure \cite{Avancini:2018svs}.} (MFIR), which uses dimensional regularization techniques, a magnetic field dependent divergence is {\it subtracted} together with other finite mass independent ($B$-dependent) terms so that, at $T=0$, one ends up with  $\Omega = \Omega_{V}(\Lambda) + \Omega_M$ where $\Omega_M$ represents a purely magnetic finite contribution that does not require further integration or sum over LL. As before, when going to finite temperatures, one needs to add the thermomagnetic contribution  $\Omega_{TM}(\Lambda)$ or  $\Omega_{TM}(\infty)$ to $\Omega$ (here we will consider the latter case which is the one most adopted in the literature). 

At this point, it is legitimate to ask how the MFIR subtraction prescription may affect physical observables since  subtracting mass-independent terms  means that  finite $B$-dependent terms also end up by being neglected. Adopting this scheme  to analyze phase transitions as in Refs. \cite {Ebert:2003yk,Menezes:2008qt} through order parameters such as the quark condensate, $\langle {\overline \psi} \psi \rangle = \partial \Omega/\partial m_c$, can be justified by the fact that  these quantities are mass dependent. However, the same scheme is certainly not  appropriate to treat quantities such as the magnetization, ${\cal M} = -\partial \Omega/\partial B$. This is because a flawless evaluation of $\cal M$  requires the knowledge of the {\it complete} $\Omega$ including all finite $B$-dependent terms  (see Ref. \cite {Endrodi:2013cs} for a related discussion). In order to circumvent this eventual problem, a fourth possibility which avoids any subtractions will be proposed in the present work. Within this prescription one starts by isolating the purely magnetic part of $\Omega$ and then identifying  two potential divergences: one  that is $B$-independent ($M$-dependent) and another one which is $B$-dependent ($M$-independent) just as in the MFIR case. However, the most important  difference is that now the $B$-dependent divergence is simply regularized but not  subtracted (by renormalizing $B^2$) as in the MFIR case. Therefore, at $T=0$, the thermodynamical potential has the form $\Omega = \Omega_{V}(\Lambda) + \Omega_M(\Lambda)$ [when considering the $T\ne0$ case in this scheme we will consider  a nonregularized piece, $\Omega_{TM}(\infty)$]. We shall 
dub this prescription vacuum magnetic regularization (VMR) scheme since  the divergences present in the vacuum  and purely magnetic contributions are regulated. 
In summary, the fact that at $B\ne0$ the Lagrangian density of nonrenormalizable models is enlarged by a {\it finite} QED type of sector  together with the additional possible choices of regularizing the convergent thermomagnetic part generates a great number of possible regularization prescriptions. Unfortunately, dealing with divergences in nonrenormalizable theories in the presence of a magnetic field and a thermal bath cannot be dealt with in a pragmatic manner as in the case of renormalizable theories where further constraints based on a well-established renormalization  programme are available. Nevertheless, one may select the most effective one by analyzing the physical behavior of different observables. With this purpose in this work, we will consider  the four possible regularization schemes already described to evaluate physical quantities such as the quark condensate, pressure, magnetization, speed of sound, and specific heat at finite temperatures and in the presence of a strong magnetic field. 
The work is organized as follows. In the next section, we present the model and review finite temperature results in the absence of magnetic fields.  Then, in Sec.III, we obtain the thermodynamical potential using  four possible regularization prescriptions. Numerical results are presented in Sec. IV and the conclusions in Sec. V.

\section{The model}
The PNJL Lagrangian density in the presence of an external magnetic field is given by \cite{Fukushima:2003fw}
\begin{equation}
    \mathcal{L}_{PNJL}=\bar{\psi}\left(i\gamma_\mu D^\mu -\hat{m}_c\right)\psi+G\left[\left(\bar{\psi}\psi\right)^2
+\left(\bar{\psi}i\gamma_5\boldsymbol\tau\psi\right)^2\right]-\mathcal{U}\left(\Phi,\bar{\Phi},T\right)-\dfrac{1}{4}F^{\mu \nu} F_{\mu \nu}, \label{PNJLlagrangian}
\end{equation}
where $\psi$ represents  fermionic fields (a sum in color and flavor indices is implicit), $\tau$ are isospin Pauli matrices, $\hat{m}_c$ are the current quark masses which, for simplicity, we set as  $m_u=m_d\equiv m_c$ while $G$ represents the coupling constant. The  covariant derivative is given  by
\begin{align}
 D^\mu=\partial^\mu-iq_fA_{EM}^\mu -iA^\mu \;,
\end{align}
where $q_f$ represents the quark electric charge,\footnote {$q_u = 2e/3, q_d=-e/3$ with $e=1/\sqrt{137}$} $A_{EM}^\mu$ is the electromagnetic gauge field, $F^{\mu \nu}=\partial^\mu A_{EM}^\mu -\partial^\nu A_{EM}^\nu$ where $A_{EM}^\mu=\delta_{\mu 2}x_1B$ and  $\vec{B}=B\hat{e}_3$ within the Landau gauge adopted here. We also consider   the Polyakov gauge where the gluonic term, $A^\mu=gA^\mu_a\left(x\right)\frac{\lambda_a}{2}$, only contributes with the spatial components:  $A^\mu=\delta^0_\mu A^0=-i\delta^0_\mu A^4$ where $g$ is the strong coupling, $A^\mu_a\left(x\right)$ represents the SU(3) gauge fields while $\lambda_a$ are the Gell-Mann matrices. \
The expectation value of the Polyakov loop, $\Phi$, is then given by the expected value of the Wilson line \cite{Fukushima:2013rx}, $L\left(\boldsymbol x\right)\equiv \mathcal{P}\exp\left[i\int^\beta_0 d\tau A_4\left(\tau,\boldsymbol x\right)\right]$. That is,
\begin{equation}
\Phi\equiv\left<\dfrac{1}{N_c}\text{Tr}L\left(\boldsymbol x\right)\right>, \ \ \ \text{and} \ \ \ \bar{\Phi}\equiv\left<\dfrac{1}{N_c}\text{Tr}L^\dagger\left(\boldsymbol x\right)\right>.
\end{equation}
Remark that  the Polyakov potential, $\mathcal{U}\left(\Phi,\bar{\Phi},T\right)$, is fixed to reproduce pure-gauge  LQCD results \cite{Ratti:2005jh}. In the case of vanishing baryonic densities ($\mu=0$) considered here, one has  $\bar\Phi=\Phi$ so that the {\it ansatz} proposed in Ref.  \cite{Ratti:2006wg} reads
\begin{equation}
\dfrac{\mathcal{U}\left(\Phi,T\right)}{T^4}=-\dfrac{1}{2}b_2\left(T\right)\Phi^2+b_4\left(T\right)\ln
\left[1-6\Phi^2+8\Phi^3-3\Phi^4\right],\label{Upot}
\end{equation}with
\begin{align}
b_2\left(T\right)=a_0+a_1\left(\dfrac{T_0}{T}\right)+a_2\left(\dfrac{T_0}{T}
\right)^2, \ \ b_4\left(T\right)=b_4\left(\dfrac
{T_0}{T}\right)^3,
\end{align}where the parametrization is given in Table \ref{tabla1}. Following Ref. \cite{Endrodi:2019whh}, we choose $T_0=208$ MeV in order to consider two quark 
flavor contributions \cite{Schaefer:2007pw}. 
\begin{table}[h!]
\begin{center}
\begin{tabular}{c c c c c} 
       
       \hline
       \hline
	$a_0$ & $a_1$ & $a_2$ & $b_4$  \\
	\hline
	3.51 &-2.47 & 15.22 & -1.75 \\
	
	\hline
	\hline
\end{tabular}
\caption[Parameter set used in the Polyakov loop potential]{Parameter set used 
for the Polyakov loop potential.} 
\label{tabla1}
\end{center}
\end{table}

\section{ Thermodynamical potential evaluations}

Let us now evaluate the thermodynamical potential, $\Omega\left(M,\Phi,T,B \right)$, by applying the MFA   to the PNJL within the PT framework. As discussed in the Introduction, the divergences will be handled in four different ways.

\subsection{TRPT  and SPT frameworks }

Within the regulated thermomagnetic integral PT formalism (TRPT) adopted in Ref. \cite{Endrodi:2019whh}, the thermodynamical potential is

\begin{eqnarray}
&&\Omega_{TRPT}(M,\Phi,T,B)=\mathcal{U}\left(\Phi,T\right)+\frac{(M-m_c)^2}{4G}+ \frac{N_c}{8\pi^2}\sum_{f=u,d}(|q_f|B)^2\int_{\frac{|q_f|B}{\Lambda^2}}^{\infty}\frac{ds}{s^2}e^{-\frac{M^2s}{|q_f|B}}\coth (s)\nonumber\\
&&+ \frac{1}{8\pi^2}\sum_{f=u,d}(|q_f|B)^2\int_{\frac{|q_f|B}{\Lambda^2}}^{\infty}\frac{ds}{s^2}e^{-\frac{M^2s}{|q_f|B}}\coth (s)\left\{ 2\sum_{n=1}^{\infty}e^{-\frac{|q_f|Bn^2}{4sT^2}}(-1)^n\left[2\cos\left( n\cos^{-1} \frac{3\Phi-1}{2}\right)+1\right]\right\} \;,
\label{OmegaPT}
\end{eqnarray}\label{therpot}
where the effective mass is given by the solution of the self-consistent gap equation
\begin{eqnarray}
&&\frac{M-m_c}{2G}=\frac{MN_c}{4\pi^2}\sum_{f=u,d}|q_f|B\int_{\frac{|q_f|B}{\Lambda^2}}^{\infty}\frac{ds}{s}e^{-\frac{M^2s}{|q_f|B}}\coth (s)\nonumber\\ 
&&+\frac{M}{4\pi^2}\sum_{f=u,d}|q_f|B\int_{\frac{|q_f|B}{\Lambda^2}}^{\infty}\frac{ds}{s}e^{-\frac{M^2s}{|q_f|B}}\coth (s)\left\{ 2\sum_{n=1}^{\infty}e^{-\frac{|q_f|Bn^2}{4sT^2}}(-1)^n\left[2\cos \left(n\cos^{-1} \frac{3\Phi-1}{2}\right)+1\right]\right\}\;,
\label{MPT}
\end{eqnarray}
\noindent which is to be solved simultaneously with  $\frac{\partial \Omega (M,\Phi,T)}{\partial \Phi}|_{M,T}=0$. 
Remark the presence of the regulator $|q_f| B/\Lambda^2$ in the lower limit of the {\it convergent} thermomagnetic integrals represented by the last terms of Eqs. (\ref {OmegaPT}) and (\ref{MPT}) which indicates that $\Omega_{TRPT}=\Omega_{VM}(\Lambda)+\Omega_{TM}(\Lambda)$. To obtain the equivalent SPT relation, one simply needs to consider these convergent terms upon performing the replacement $|q_f| B/\Lambda^2 \to 0$ in the lower limit of those integrals so that $\Omega_{SPT}=\Omega_{VM}(\Lambda)+\Omega_{TM}(\infty)$ as already discussed.
For completeness, let us also quote the $B=0$ relations,
\begin{eqnarray}
\Omega_{TRPT}(M,\Phi,T,0)=\mathcal{U}\left(\Phi,T\right)+\frac{(M-m_c)^2}{4G}+\frac{N_cN_f}{8\pi^2}\int_{\frac{1}{\Lambda^2}}^{\infty}\frac{ds}{s^3}e^{-M^2s}+\nonumber\\\frac{N_f}{8\pi^2}\int_{\frac{1}{\Lambda^2}}^{\infty}\frac{ds}{s^3}e^{-M^2s}\left\{ 2\sum_{n=1}^{\infty}e^{-\frac{n^2}{4sT^2}}(-1)^n\left[2\cos \left(n\cos^{-1} \frac{3\Phi-1}{2}\right)+1\right]\right\}
\label{Omega0}
\end{eqnarray}
and

\begin{equation}
\frac{M-m_c}{2G}=\frac{MN_cN_f}{4\pi^2}\int_{\frac{1}{\Lambda^2}}^{\infty}\frac{ds}{s^2}e^{-M^2s}+\frac{MN_f}{4\pi^2}\int_{\frac{1}{\Lambda^2}}^{\infty}\frac{ds}{s^2}e^{-M^2s}\left\{ 2\sum_{n=1}^{\infty}e^{-\frac{n^2}{4sT^2}}(-1)^n\left[2\cos \left(n\cos^{-1} \frac{3\Phi-1}{2}\right)+1\right]\right\}.
\label{gap0}
\end{equation}
\noindent 
To obtain the equivalent VMR relations one performs the replacement  $1/\Lambda^2 \to 0$ in the lower limit of the convergent thermal integrals represented by the last terms of Eqs. (\ref{Omega0}) and (\ref{gap0}). 

\subsection{ VMR and MFIR frameworks}

The first step to implement the vacuum magnetic regularization scheme proposed here is to split the (divergent) third term of Eq. (\ref{OmegaPT}) in one $B$-independent integral and one pure magnetic expression (see Appendix \ref{AppendixA} for details). Then,
\begin{equation}\label{split1}
\begin{split}
 \frac{N_c}{8\pi^2}\sum_{q_f=u,d}(|q_f|B)^2\int_{\frac{|q_f|B}{\Lambda^2}}^{\infty}\frac{ds}{s^2}e^{-\frac{M^2s}{|q_f|B}}\coth (s)= &\frac{N_cN_f}{8\pi^2}\int_{\frac{1}{\Lambda^2}}^{\infty}\frac{ds}{s^3}e^{-M^2s} +\frac{N_c}{24\pi^2}\sum_{q_f=u,d}(|q_f|B)^2\left[\ln\left(\frac{\Lambda^2}{2|q_f|B}\right)+1-\gamma_{E}\right]\\
 &-N_c\sum_{f=u,d}\dfrac{(|q_f|B)^2}{2\pi^2}\left[\zeta^\prime\left(-1,x_f\right)-\left[x_f^2-x_f\right]\frac{\ln x_f}{2}+\dfrac{x_f^2}{4}\right],
 \end{split}
\end{equation}
where $x_f=\frac{M^2}{2|q_f|B}$, and $\zeta$ represents the Hurwitz-Riemann-zeta function.

Then, considering Eq. (\ref{split1}) together with the finite thermomagnetic contribution, one obtains the full VMR thermodynamical potential
\begin{eqnarray}
&&\Omega_{VMR} (M,\Phi,T,B)=\mathcal{U}\left(\Phi,T\right)+\frac{(M-m_c)^2}{4G}+\frac{N_cN_f}{8\pi^2}\int_{\frac{1}{\Lambda^2}}^{\infty}\frac{ds}{s^3}e^{-M^2s} +\frac{N_c}{24\pi^2}\sum_{q_f=u,d}(|q_f|B)^2\left[\ln\left(\frac{\Lambda^2}{2|q_f|B}\right)+1-\gamma_{E}\right]\nonumber\\
 &&-N_c\sum_{f=u,d}\dfrac{(|q_f|B)^2}{2\pi^2}\left[\zeta^\prime\left(-1,x_f\right)-\dfrac{1}{2}\left[x_f^2-x_f\right]\ln x_f+\dfrac{x_f^2}{4}\right]\nonumber\\
&&+ \frac{1}{8\pi^2}\sum_{f=u,d}(|q_f|B)^2\int_{0}^{\infty}\frac{ds}{s^2}e^{-\frac{M^2s}{|q_f|B}}\coth (s)\left\{ 2\sum_{n=1}^{\infty}e^{-\frac{|q_f|Bn^2}{4sT^2}}(-1)^n\left[2\cos\left( n\cos^{-1} \frac{3\Phi-1}{2}\right)+1\right]\right\} \;,
\label{OmegaVMR}
\end{eqnarray}
which is clearly of the form $\Omega_{VMR}=\Omega_V(\Lambda)+ \Omega_{M}(\Lambda)+\Omega_{TM}(\infty)$.
For completeness, let us recall  the  equivalent MFIR equation \cite {Ebert:2003yk,Menezes:2008qt,Avancini:2019wed,Allen:2015paa}, 
\begin{eqnarray}
&&\Omega_{MFIR} (M,\Phi,T,B)=\mathcal{U}\left(\Phi,T\right)+\frac{(M-m_c)^2}{4G}+\frac{N_cN_f}{8\pi^2}\int_{\frac{1}{\Lambda^2}}^{\infty}\frac{ds}{s^3}e^{-M^2s}\nonumber\\
 &&-N_c\sum_{f=u,d}\dfrac{(|q_f|B)^2}{2\pi^2}\left[\zeta^\prime\left(-1,x_f\right)-\dfrac{1}{2}\left[x_f^2-x_f\right]\ln x_f+\dfrac{x_f^2}{4}\right]\nonumber\\
&&+ \frac{1}{8\pi^2}\sum_{f=u,d}(|q_f|B)^2\int_{0}^{\infty}\frac{ds}{s^2}e^{-\frac{M^2s}{|q_f|B}}\coth (s)\left\{ 2\sum_{n=1}^{\infty}e^{-\frac{|q_f|Bn^2}{4sT^2}}(-1)^n\left[2\cos\left( n\cos^{-1} \frac{3\Phi-1}{2}\right)+1\right]\right\} \;,
\label{OmegaMFIR}
\end{eqnarray}
which is  of the form $\Omega_{MFIR}=\Omega_V(\Lambda)+ \Omega_{M} +\Omega_{TM}(\infty)$. Therefore, one can relate the VMR and MFIR results as 
\begin{equation}
\Omega_{VMR} (M,\Phi,T,B)= \Omega_{MFIR} (M,\Phi,T,B) +\frac{N_c}{24\pi^2}\sum_{q_f=u,d}(|q_f|B)^2\left[\ln\left(\frac{\Lambda^2}{2|q_f|B}\right)+1-\gamma_{E}\right] \,,
\label{compare}
\end{equation}
which clearly displays the most important difference between the two regularization prescriptions. Namely, within the MFIR, the divergence of the magnetic contribution represented by the  $\ln[\Lambda^2/(2 |q_f|B)]$ is completely subtracted by renormalizing $B^2$. However, in this process, important finite contributions represented by the terms proportional to  $(1-\gamma_E)$ are also canceled directly impacting the magnetization. 
Finally, the VMR gap equation (which is exactly the same as the MFIR one) is given by
\begin{eqnarray}
&&\frac{M-m_c}{2G}=\frac{MN_cN_f}{4\pi^2}\int_{\frac{1}{\Lambda^2}}^{\infty}\frac{ds}{s^2}e^{-M^2s}+M N_c\sum_{f=u,d}\dfrac{|q_f B|}{2\pi^2}
\left[\ln\left(\Gamma\left[x_f\right]\right)-\dfrac{1}{2}
\ln\left(2\pi\right)+x_f-\dfrac{1}{2}\left(2x_f-1\right)\ln x_f\right]\nonumber\\
&&+\frac{M}{4\pi^2}\sum_{f=u,d}|q_f|B\int_{0}^{\infty}\frac{ds}{s}e^{-\frac{M^2s}{|q_f|B}}\coth (s)\left\{ 2\sum_{n=1}^{\infty}e^{-\frac{|q_f|Bn^2}{4sT^2}}(-1)^n\left[2\cos \left(n\cos^{-1} \frac{3\Phi-1}{2}\right)+1\right]\right\}\;.
\label{MVMR}
\end{eqnarray}
As one can see, the VMR extra (mass independent) term contained in Eq. (\ref {compare}) does not contribute to the mass gap nor to higher derivatives with respect to the mass and thus observes Goldstone's theorem \cite{Gamayun:2012qr,Cao:2014uva}.

\section {Results}

Having the thermodynamical potential, we can easily obtain some important thermodynamical observables such as the  pressure, $P=-\Omega$, and  the energy density, ${\cal E} = - P + T {\cal S} + B {\cal M}$, where the entropy density is  ${\cal S}= \partial P / \partial T$ and the magnetization is ${\cal M}= \partial P / \partial B$. 
For our purposes, it will prove useful to also investigate the  quark condensate
\begin{equation}
\langle {\overline \psi_f} \psi_f \rangle = - \frac{\partial P}{\partial m_c} \,,
\end{equation}
as well as the specific heat and the speed of sound squared which are, respectively, defined as
\begin{equation}
C_v = T \frac {\partial {\cal S}}{\partial T}\;
\end{equation}
and 
 \begin{equation}
C_s^2 = \frac{\partial P}{\partial {\cal E}} = \frac{\cal S} {C_v}\,.
\end{equation}
As explained in the Introduction, in order to be in line with LQCD predictions, we shall consider a $B$-dependent coupling, $G(B)$, whose running was determined in Ref. \cite{Endrodi:2019whh}. Following that work, we first set $\Lambda=675$ MeV and $m_c=3.5$ MeV and then tune $G(B)$  so that all four different regularization schemes considered here yield the quark mass value needed to reproduce the mesonic masses predicted by LQCD simulations.
 The values of the dimensionless quantity $G(B) \Lambda^2$ at  different   magnetic intensities are given in Table \ref{tabla2} for the four different schemes.\\
 \begin{figure}
 \center
\subfigure{\includegraphics[width=0.4\textwidth]{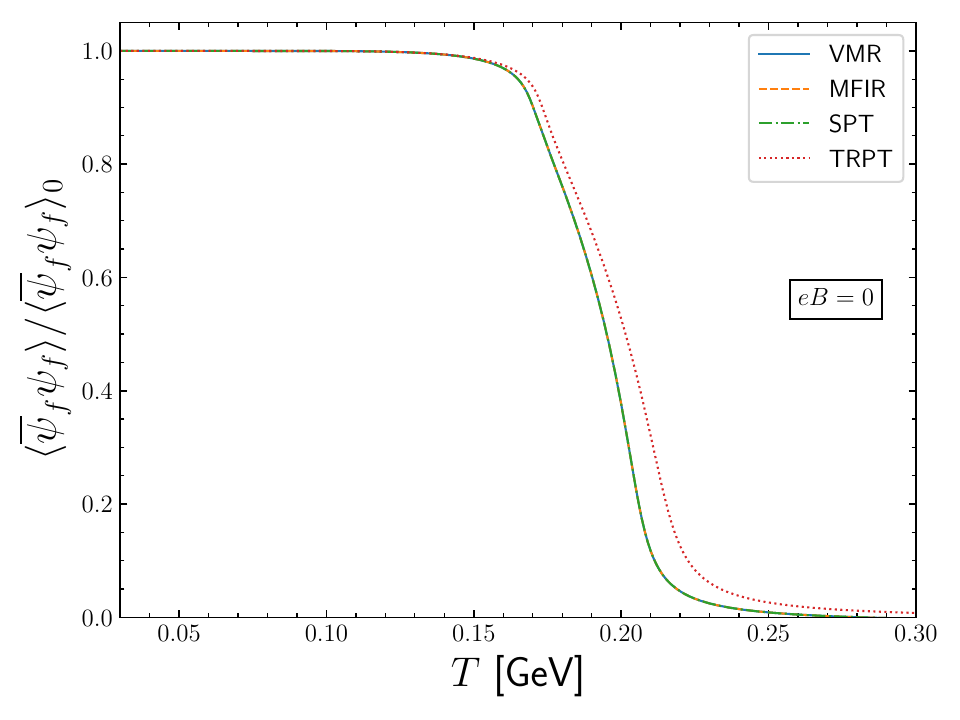}}
\subfigure{\includegraphics[width=0.4\textwidth]{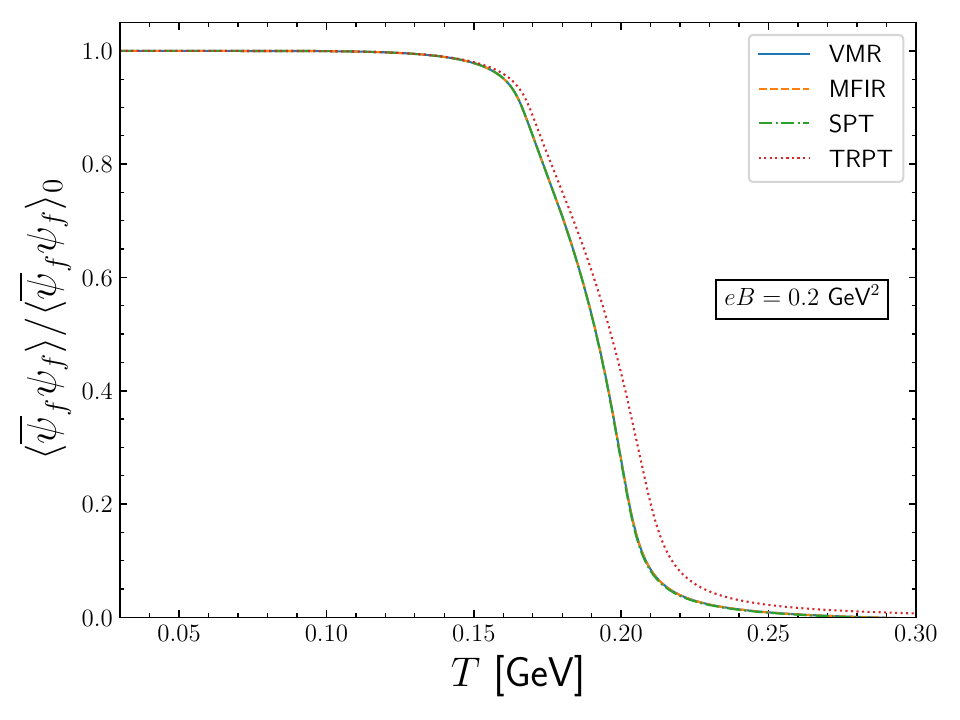}}
\subfigure{\includegraphics[width=0.4\textwidth]{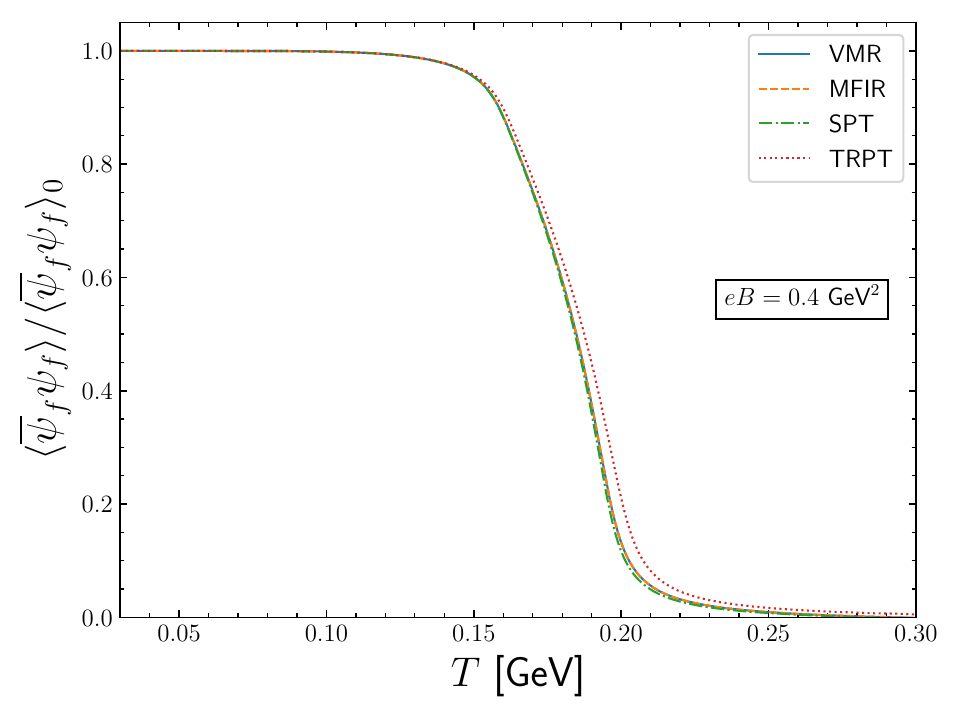}}
\subfigure{\includegraphics[width=0.4\textwidth]{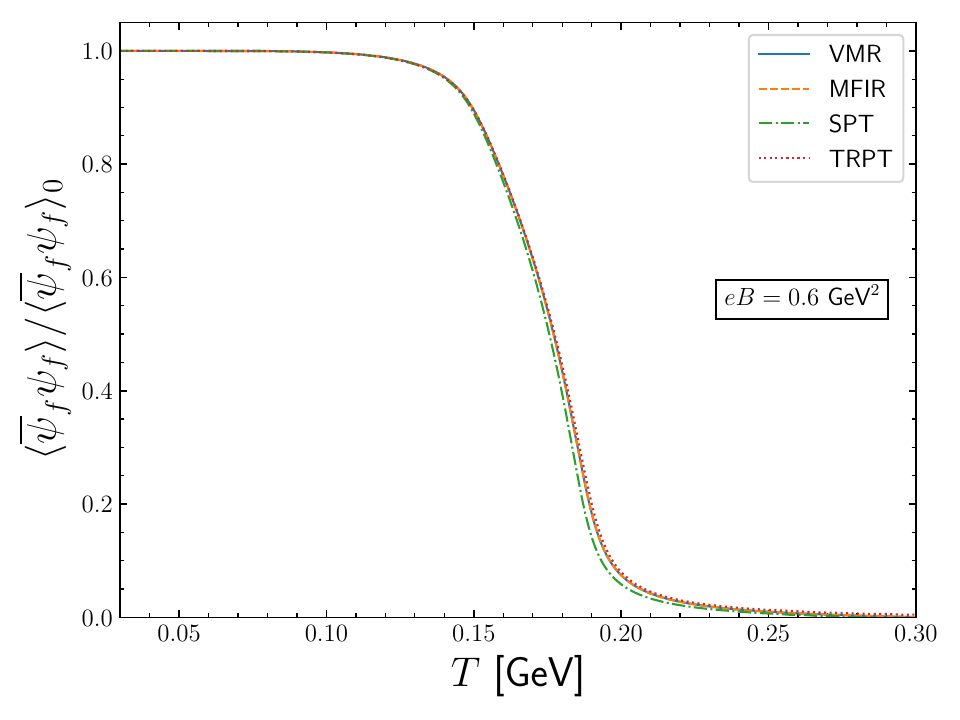}}
 \caption{Normalized quark condensate for different magnetic fields as a function of temperature calculated with the different regularization procedures. The quantity $\langle\bar{\psi}_f\psi_f\rangle_0$ represents the quark condensate at $T=0$.}
  \label{cond}
\end{figure}
\begin{figure}
  \subfigure{\includegraphics[width=0.4\textwidth]{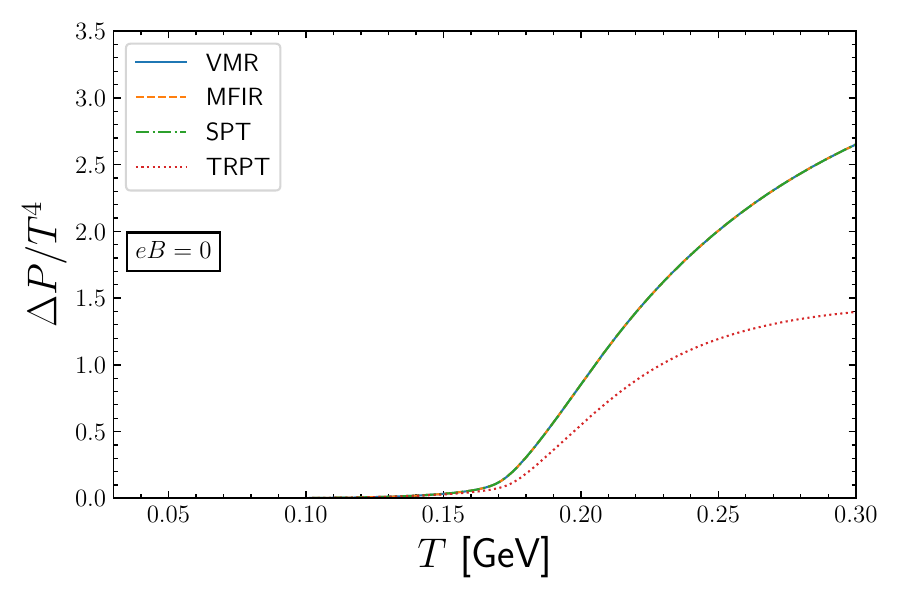}}
  \subfigure{\includegraphics[width=0.4\textwidth]{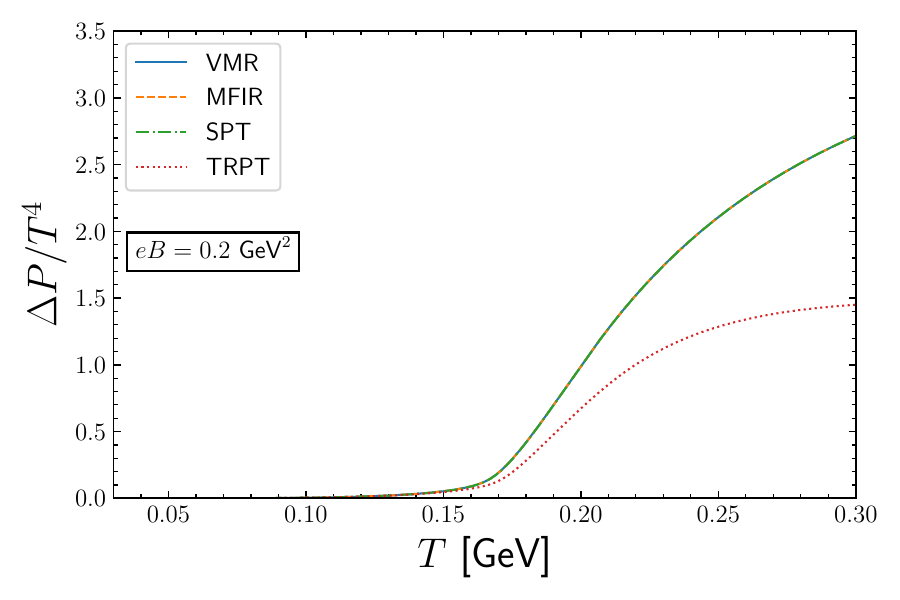}}
  \subfigure{\includegraphics[width=0.4\textwidth]{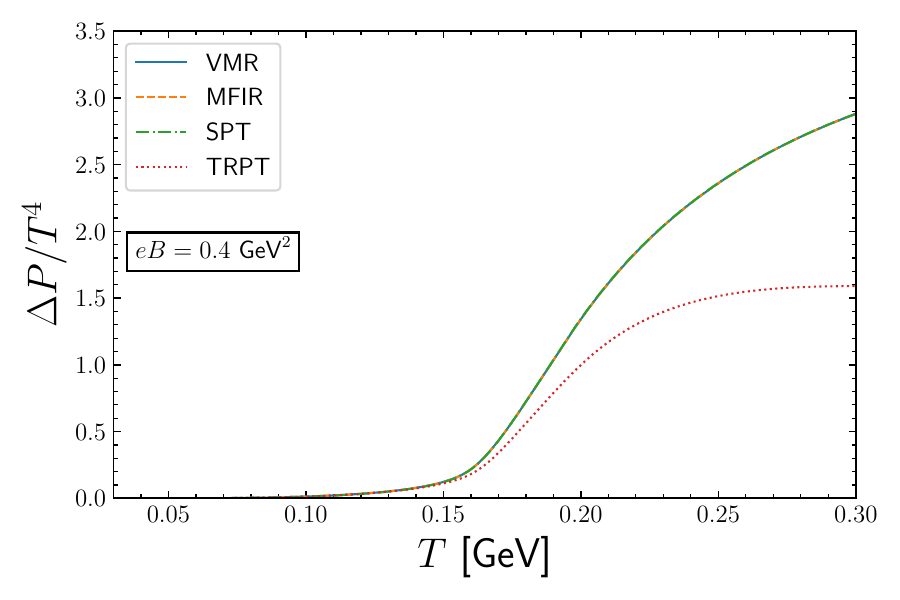}}
  \subfigure{\includegraphics[width=0.4\textwidth]{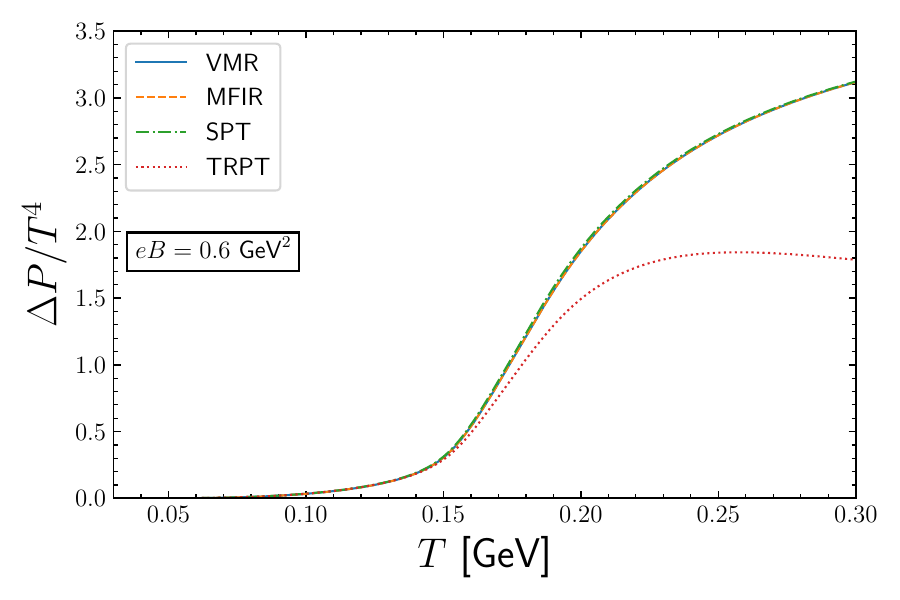}}
 \caption{Normalized pressure for different magnetic field values as a function of temperature calculated with the different regularization procedures. }
  \label{Npressure}
\end{figure}
\begin{figure}
  \subfigure{\includegraphics[width=0.4\textwidth]{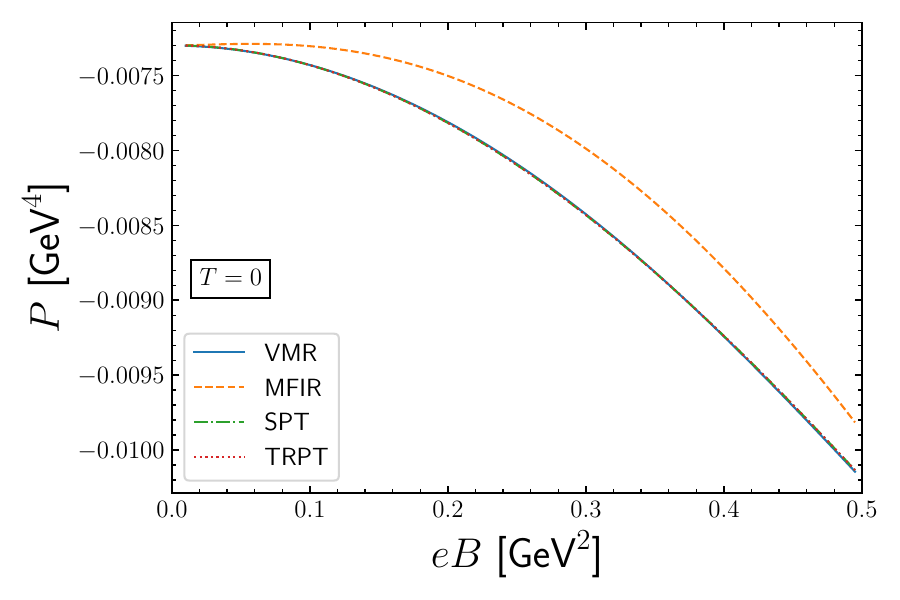}}
  \subfigure{\includegraphics[width=0.4\textwidth]{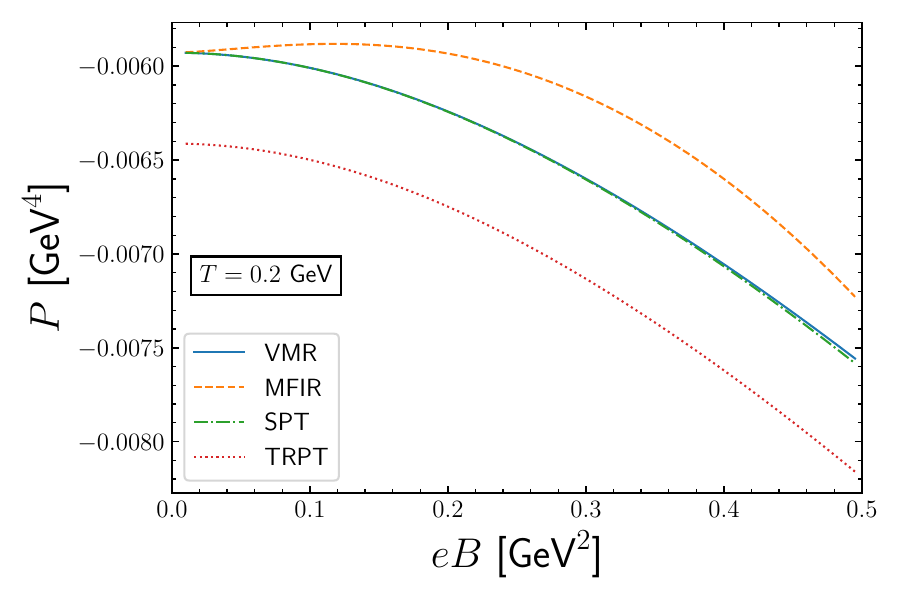}}
  \subfigure{\includegraphics[width=0.4\textwidth]{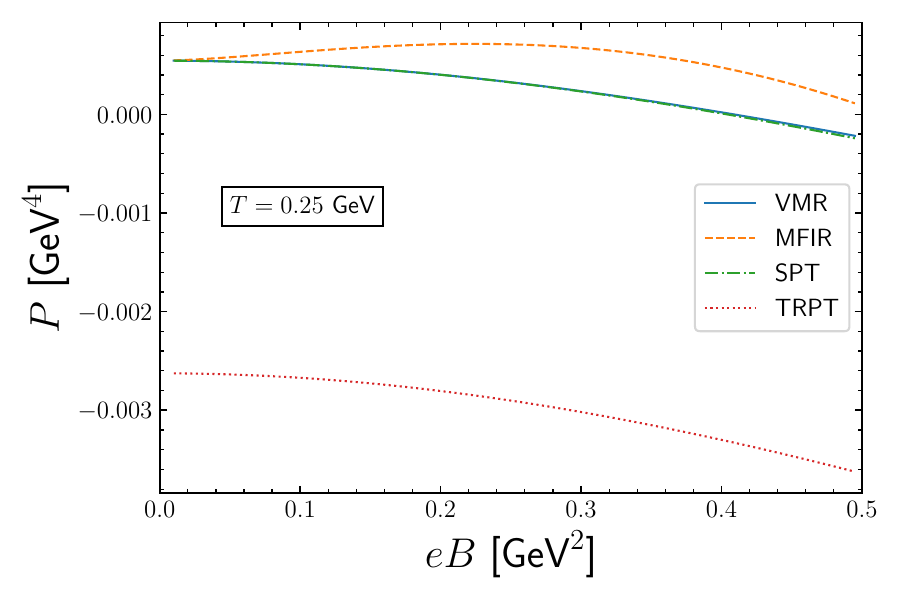}}
 \caption{Pressure for different temperature values as a function of $eB$ calculated with the different regularization procedures. }
  \label{pressure}
\end{figure}
\begin{figure}
\subfigure{\includegraphics[width=0.4\textwidth]{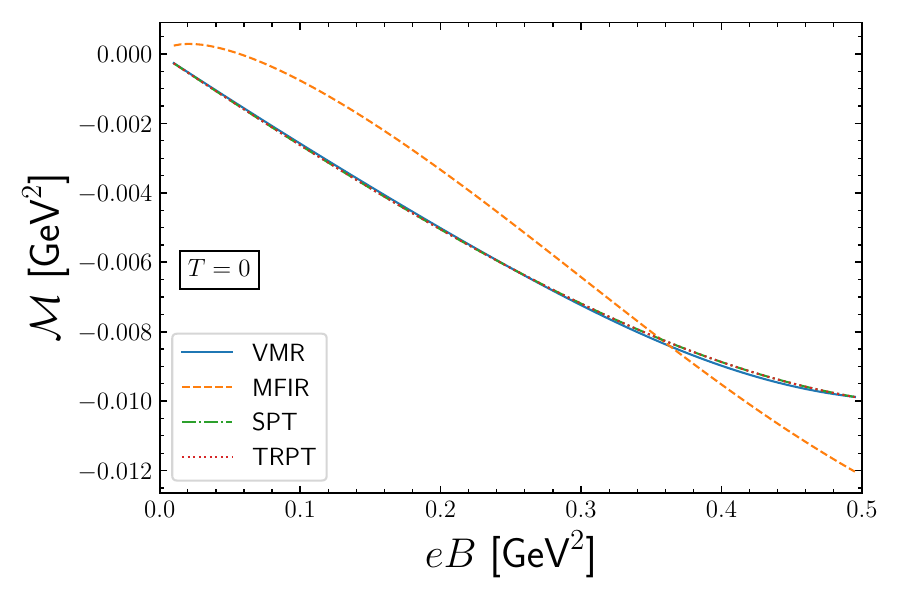}}
\subfigure{\includegraphics[width=0.4\textwidth]{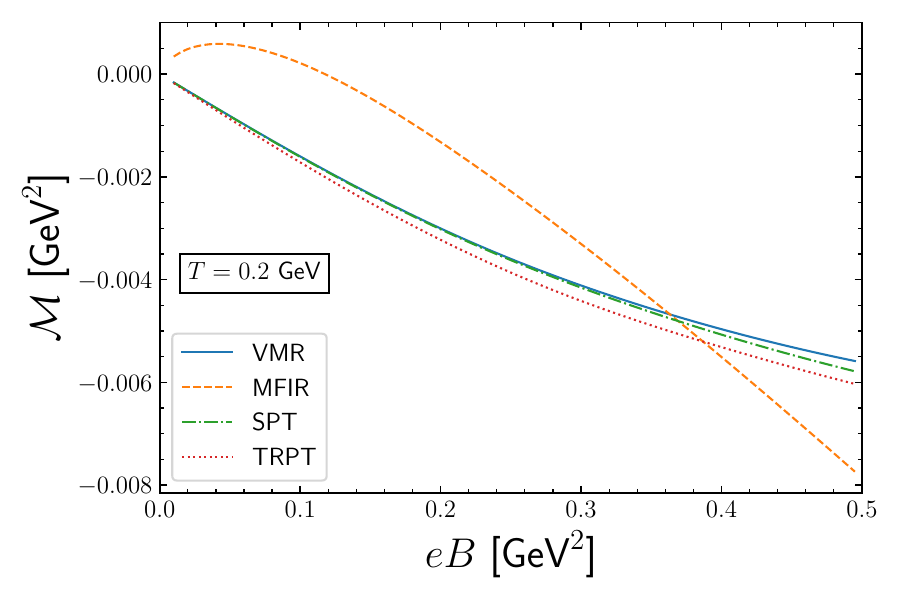}}
\subfigure{\includegraphics[width=0.4\textwidth]{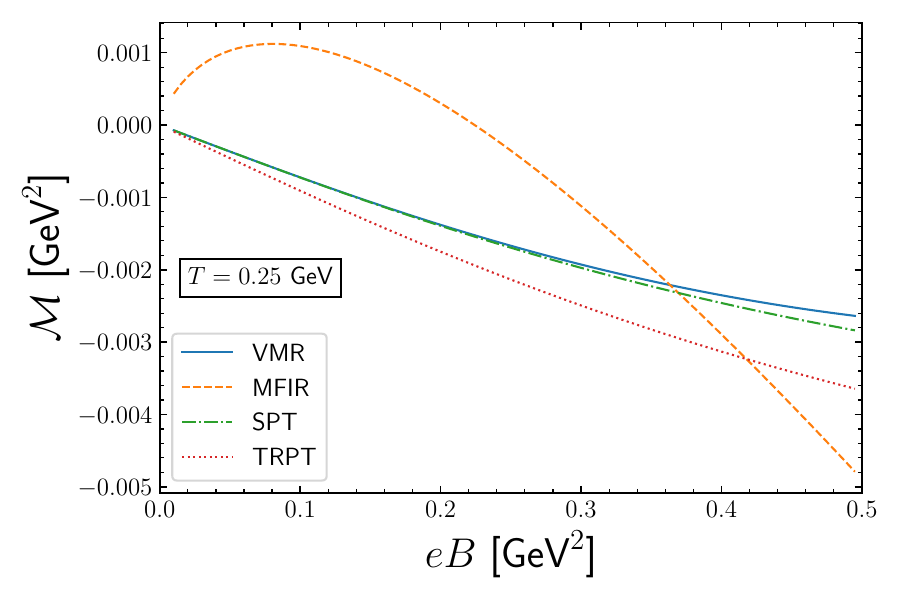}}
 \caption{Magnetization for different temperature values  as a function of $eB$ calculated with the different regularization procedures.  }
  \label{MagvsB}
\end{figure}
\begin{figure}
\subfigure{\includegraphics[width=0.4\textwidth]{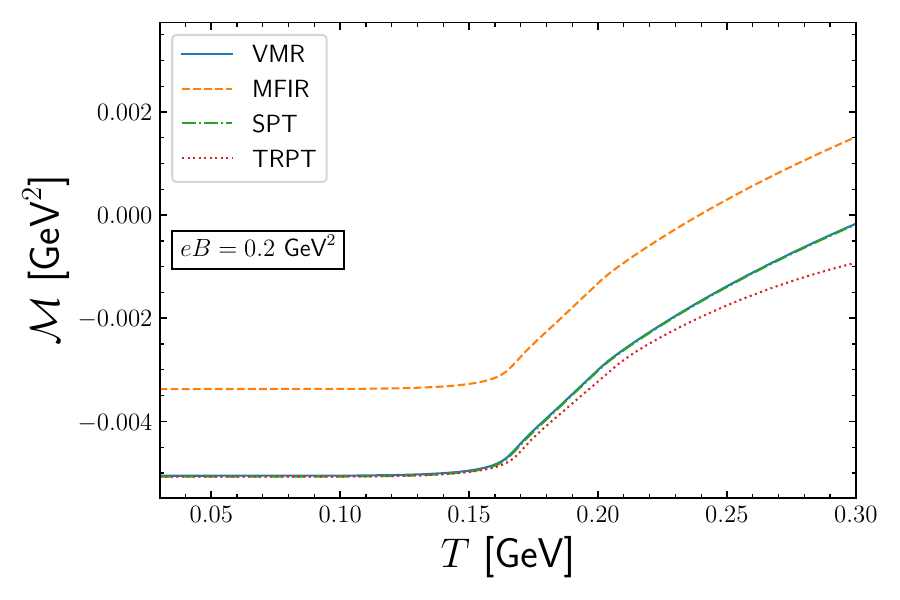}}
\subfigure{\includegraphics[width=0.4\textwidth]{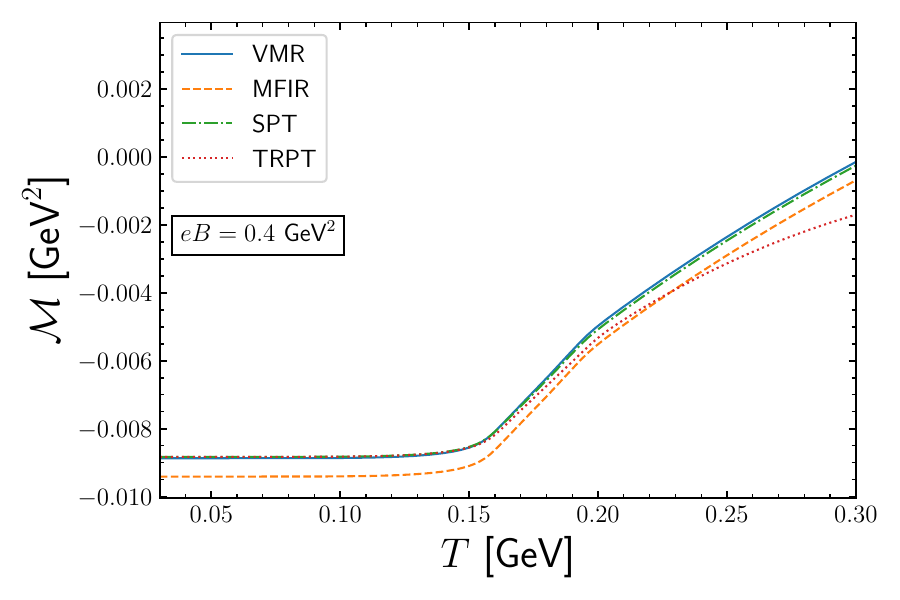}}
\subfigure{\includegraphics[width=0.4\textwidth]{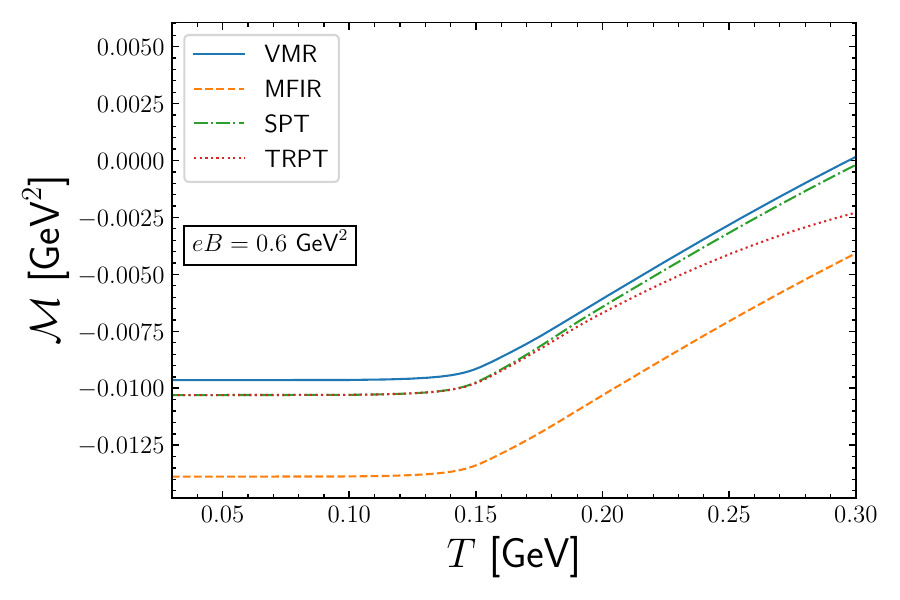}}
 \caption{Magnetization for different magnetic field values  as a function of the temperature calculated with the different regularization procedures. }
  \label{MagvsT}
\end{figure}
\begin{figure}
\subfigure{\includegraphics[width=0.4\textwidth]{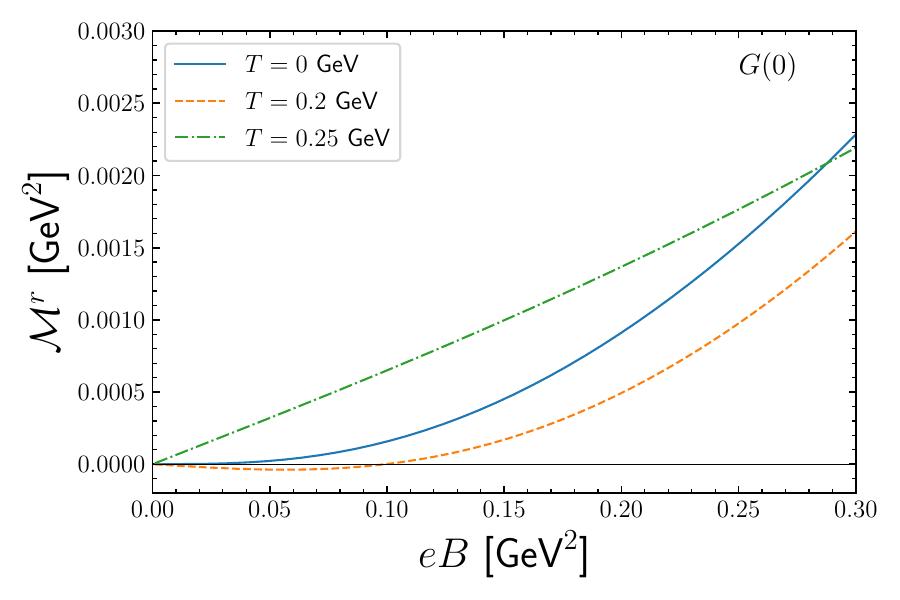}}
\subfigure{\includegraphics[width=0.4\textwidth]{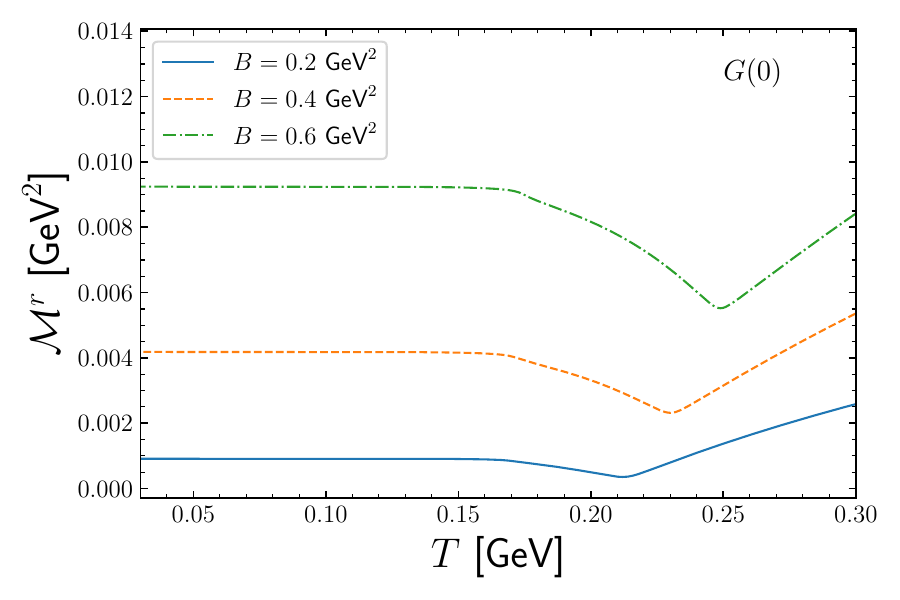}}

 \caption{Left panel: renormalized magnetization for a fixed coupling, $G(0)$, at different temperatures  as a function of the magnetic field. 
 Right panel: renormalized magnetization for a fixed coupling, $G(0)$, at different magnetic field values  as a function of the temperature. }
  \label{MrvsTB}
\end{figure}
\begin{figure}
  \subfigure{\includegraphics[width=0.4\textwidth]{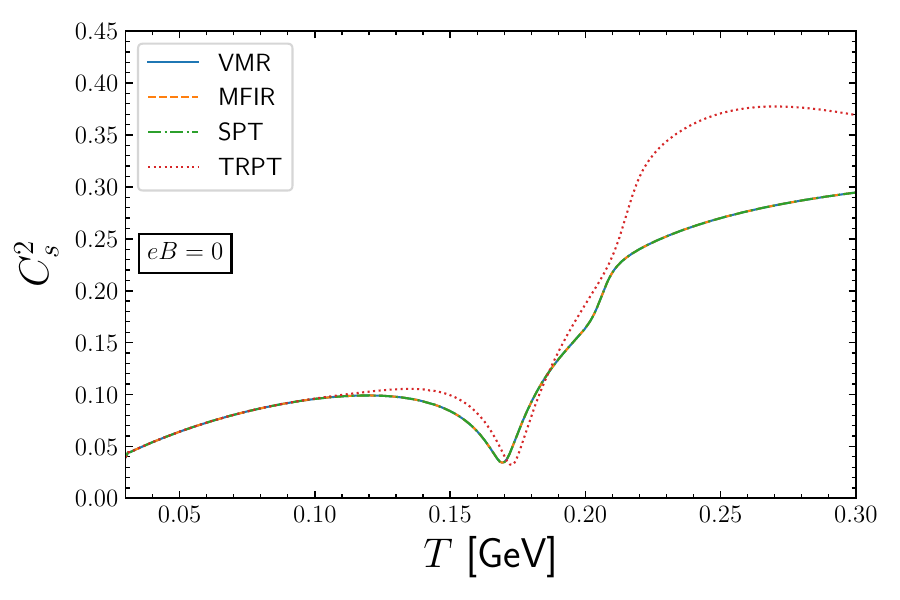}}
   \subfigure{\includegraphics[width=0.4\textwidth]{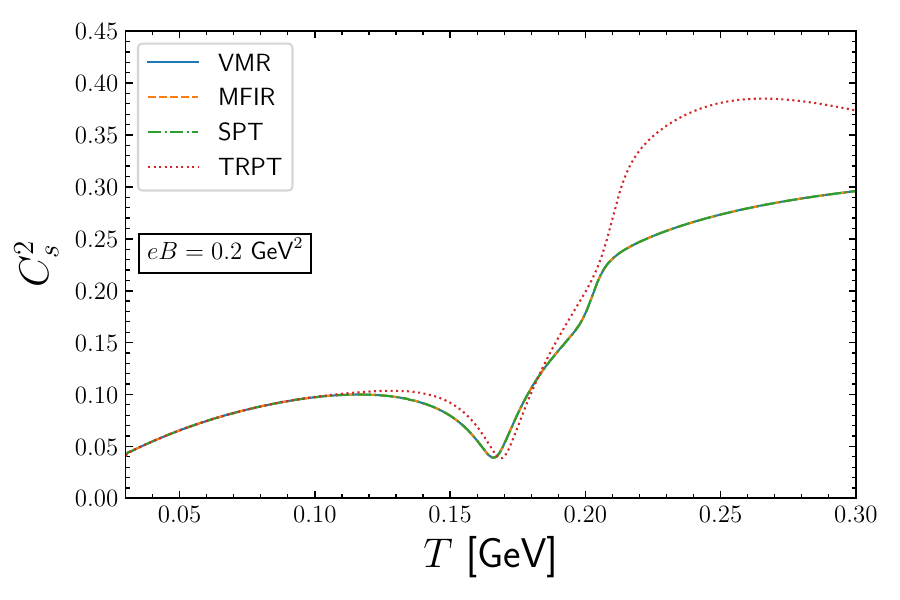}}
    \subfigure{\includegraphics[width=0.4\textwidth]{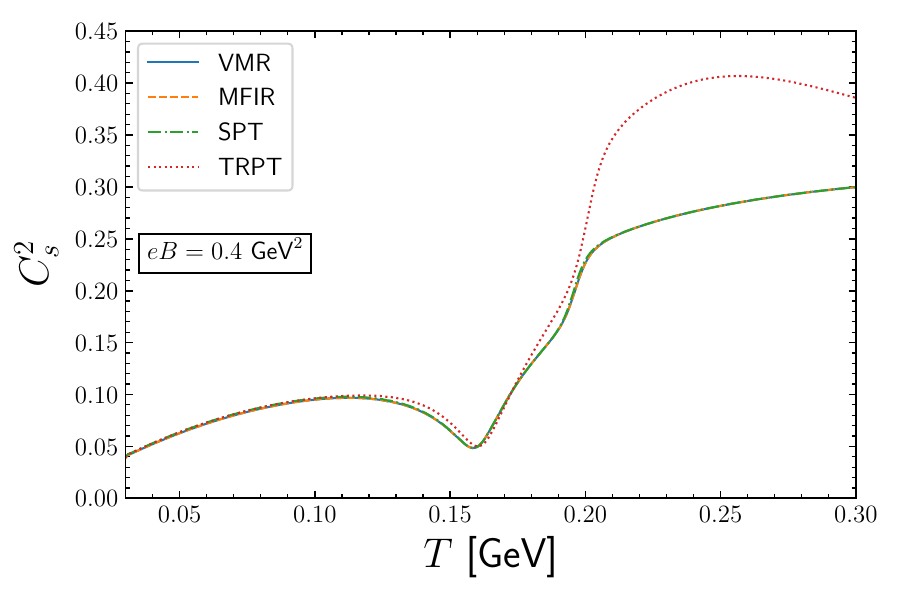}}
    \subfigure{\includegraphics[width=0.4\textwidth]{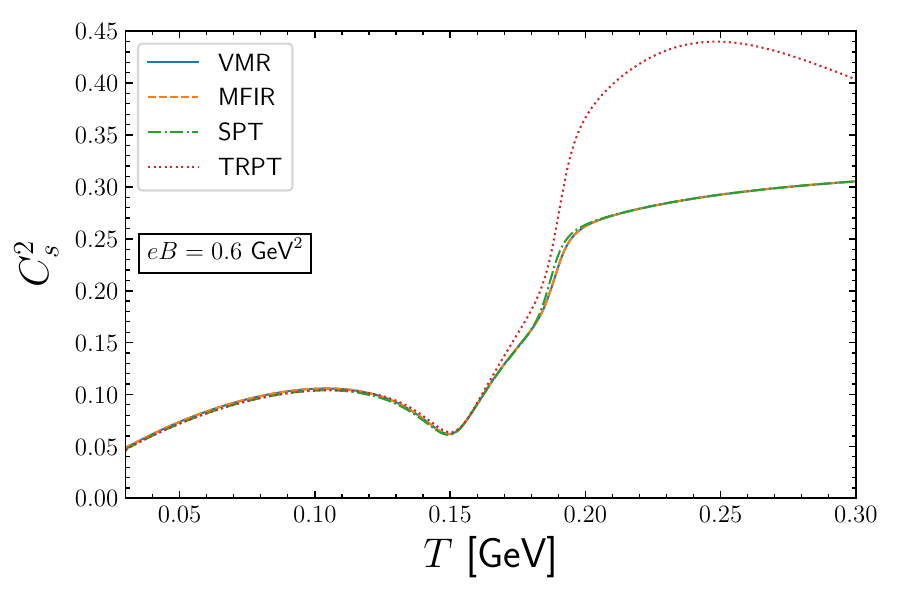}} 
 \caption{The squared speed of sound for different magnetic field values as a function of temperature calculated with the different regularization procedures }
  \label{Cs}
\end{figure}
\begin{figure}
  \subfigure{\includegraphics[width=0.45\textwidth]{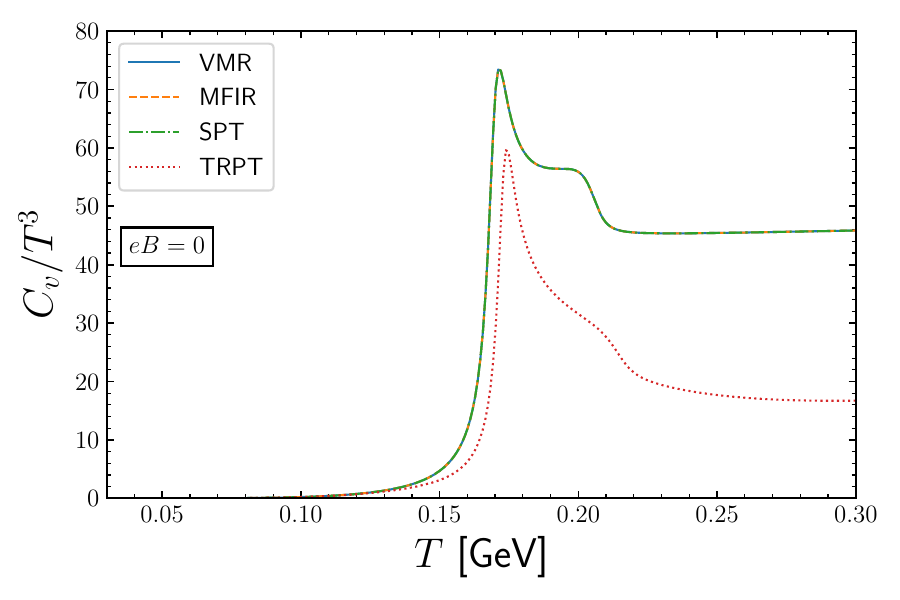}}
   \subfigure{\includegraphics[width=0.45\textwidth]{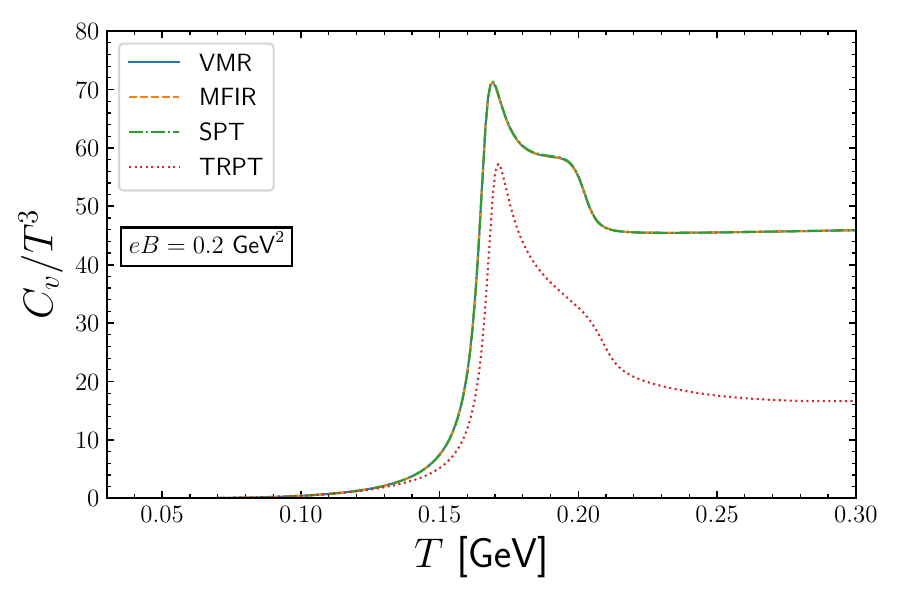}}
   \subfigure{\includegraphics[width=0.45\textwidth]{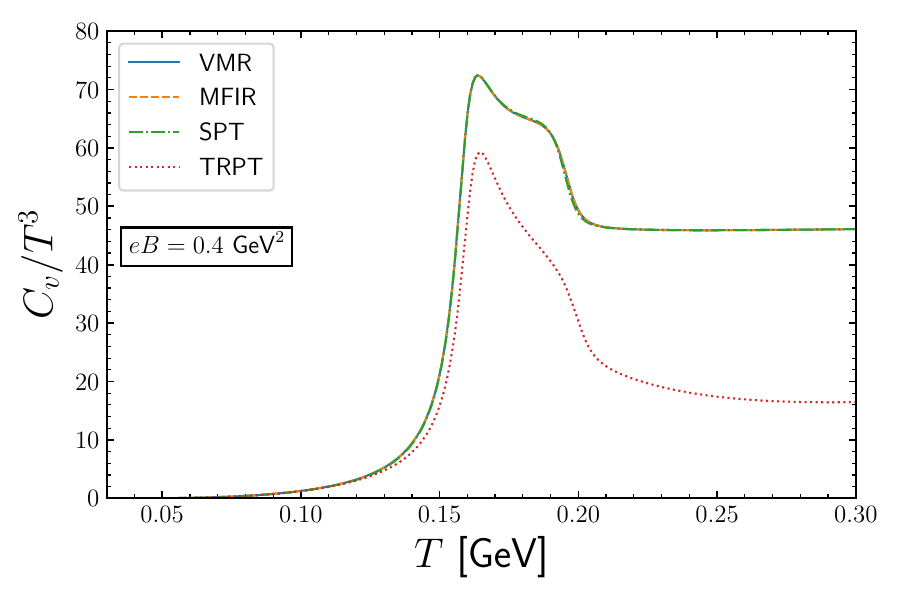}}
   \subfigure{\includegraphics[width=0.45\textwidth]{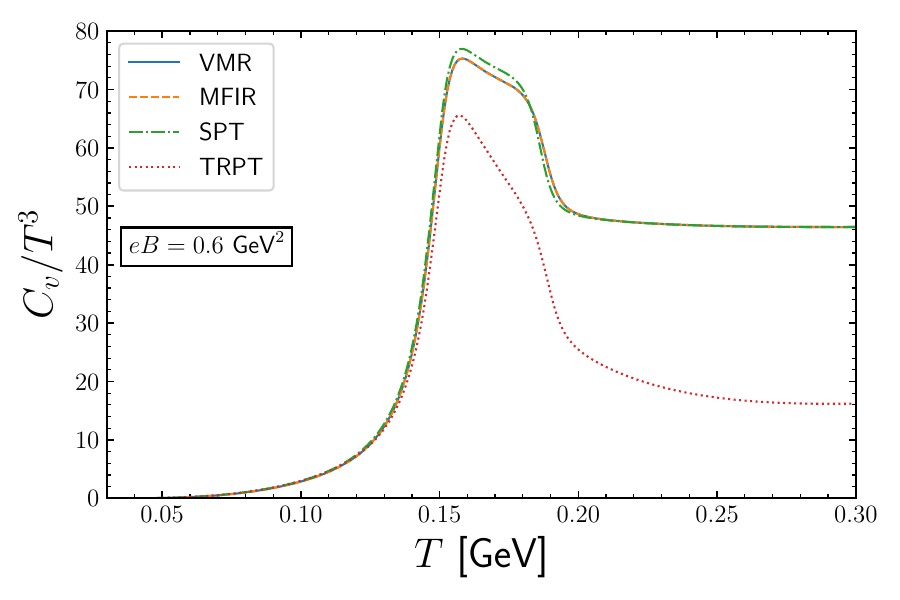}} 
 \caption{The specific heat for different magnetic field values as a function of temperature calculated with the different regularization procedures.  }
  \label{Cv}
\end{figure}
\begin{table}[h!]
\begin{center}
\begin{tabular}{|c |c| c| } 
       \hline
	 $eB$ [$\text{GeV}^2$] &VMR and MFIR  &SPT {\rm and} TRPT\\
	\hline
	\hline
	0.0 &  5.83200 & 5.83200\\
	0.2 & 5.05349  & 5.19413 \\
	0.4  & 3.74477  & 4.05506 \\
	0.6  & 2.69719  & 3.05269 \\
	\hline	
\end{tabular}
\caption{$G(B) \Lambda^2$ values for the four different regularization prescriptions. } 
\label{tabla2}
\end{center}
\end{table}
Let us start by analyzing the chiral transition order parameter represented by the quark condensate. Figure \ref {cond} shows this quantity as a function of the temperature for different values of the magnetic field illustrating that except for the TRPT all other regularization schemes predict a similar quantitative behavior. As $B$ increases, the TRPT predictions are in better agreement with the ones furnished by the other three prescriptions. It is also clear, from the inflection points, that the pseudocritical temperature value decreases as $B$ increases as one could anticipate. The subtracted pressure, $\Delta P= P(T,B)-P(0,B)$, as a function of $T$ is presented in Fig. \ref{Npressure} for different values of $B$. One can now observe that the TRPT scheme also produces a rather different high-$T$ behavior which is enhanced as higher magnetic fields are considered as the panel for the $eB=0.6 \; {\rm GeV}^2$ suggests. The maximum at $T \approx  0.23 \; {\rm GeV}$ ($eB=0.6 \; {\rm GeV}^2$) is a reminder that by regulating the (convergent) thermomagnetic integrals one loses predictive power at high temperatures. As already emphasized, a major drawback of this type of regularization procedure is that the Stefan-Boltzmann limit is never attained as $T\to \infty$ \cite {Zhuang:1994dw}. Next, to illustrate the effect of the missing mass-independent terms in the MFIR as well as the effect of regulating the thermomagnetic integrals within the TRPT schemes, we offer Fig. \ref {pressure}. This figure shows that  the TRPT predictions become less reliable as $T$ increases,  in accordance with our previous discussion. Moreover, the figure clearly illustrates  how the neglected mass-independent terms seem to affect the pressure by causing its absolute value to first decrease (at low $B$) and then increase after having reached an extremum. This behavior is in contradiction with the ones predicted by all the other three schemes  and directly affects the magnetization as Fig. \ref{MagvsB} shows. From the qualitative point of view, it is important to note that the MFIR predicts quark matter to be  paramagnetic (${\cal M}>0$) at low $B$ and diamagnetic (${\cal M}<0$) at high magnetic fields, while the other approximations predict it to be diamagnetic (at least up to the highest temperature considered in the figure, $T=0.25 \; {\rm GeV}$). The thermal behavior of the magnetization can be better analyzed by plotting this quantity as a function of $T$ for different values of $B$ as Fig. \ref{MagvsT} shows.  One can observe that the thermal behavior predicted by the MFIR is very sensitive to variations of $B$. At $eB=0.2\; {\rm GeV}^2$, the predicted MFIR absolute value for $\cal M$ is lesser than the ones predicted by the other approximations. Then, all the predicted values almost coincide at $eB=0.4\; {\rm GeV}^2$ while the MFIR  absolute values are higher at $eB=0.6\; {\rm GeV}^2$.

At this point, a digression concerning the magnetic character of the QCD vacuum is in order since  LQCD evaluations, at $T=0$, have shown that the vacuum is paramagnetic \cite {Bali:2013esa} in contradiction to our present findings. 
The VMR can indeed be reconciled with the LQCD results 
provided that one uses the {\it same} definition for the {\it renormalized} magnetization, ${\cal M}^r$, used in Ref. \cite {Bali:2013esa} where the authors consider
\begin{equation}
{\cal M}^r \cdot eB = {\cal M} \cdot eB - \left.(eB)^2 \displaystyle{\lim_{eB \to 0}} \frac{{\cal M} \cdot eB}{(eB)^2}\right|_{T=0} \,.
\label{latticeREN}
\end{equation}
In Appendix \ref{AppendixB}, we  give the analytical details necessary to obtain this quantity within our framework. 
Figure \ref{MrvsTB} shows ${\cal M}^r$ varying with the magnetic field (left panel) and with the temperature (right panel) when considering the coupling fixed at the $B=0$ value, $G(0)$. Note that although a fixed coupling does not favor inverse magnetic catalysis, the  paramagnetic behavior dominates most regions  as the left panel shows.  On the other hand,  our preliminary investigations suggest that  addressing this issue in an appropriate manner will require the determination of how $G$  runs with $B$ with much greater accuracy than the  one provided by  Ref. \cite {Endrodi:2019whh} (which was adopted here). 
This nontrivial research programme, which may also  require the consideration 
of strangeness,\footnote{Recall that the lattice results  \cite {Bali:2013esa} where obtained at $N_f=2+1$ while here only the $N_f=2$ case is considered.}  
demands a careful numerical analysis.  Therefore, we postpone this investigation for  a future application since for our present purposes the results shown in Fig. \ref{MrvsTB} 
should convince the reader that the apparent qualitative disagreement between the results shown in Figs \ref {MagvsB} and \ref {MagvsT} and lattice predictions can indeed be 
reconciled. For that, one needs to consider the renormalized magnetization, ${\cal M}^r$, in conjunction with  a coupling which runs with $B$ in a  way which is different from 
the one proposed in Ref. \cite {Endrodi:2019whh}. This last remark is corroborated by the fact that in the extreme  $G(0)$ case (no running at all) our VMR predicts quark matter to be paramagnetic in accordance with lattice simulations. Also, when implementing such a programme, one should  consider the quark condensates in the vacuum to be    $\left \langle \bar{u}u \right \rangle+\left \langle \bar{d}d \right \rangle=-2(0.25 \text{GeV})^3$ \cite{Klevansky:1992qe}.

It is important to emphasize that the  VMR scheme proposed here is 
not totally compatible with the renormalized magnetization. This can be understood by recalling that the purely magnetic contribution guarantees the magnetic 
catalysis of the VMR gap equation. However, in renormalizable theories such as QED considered by  Schwinger  \cite{Schwinger:1951nm} or QCD considered in LQCD 
applications \cite{Bali:2013esa}, one expects to renormalize 
the additional $\mathcal{O}(eB)^2$ contribution through the term $B^2/2$. Therefore, the renormalized magnetization on the PNJL SU(2) model with VMR regularization discussed in Appendix \ref{AppendixB}
is merely a projected magnetization that can be a useful tool as far as  comparisons with LQCD are concerned. In other words, by  doing this, we hope to have a magnetization as similar as possible to that considered in
lattice applications  \cite{Bali:2013esa}.  

Coming back to Fig. \ref {MagvsT} another interesting feature concerns the high-$T$ behavior of the TRPT curves which increase with $T$ in a less linear fashion than all other curves. This could be anticipated based on our previous discussion related to the Stefan-Boltzmann limit. Such a problem becomes  even more transparent when one analyzes the speed of sound squared as a function of the temperature. In this case, Fig. \ref {Cs}  clearly reveals that for $ T \gtrsim 0.2\,{\rm GeV}$ the TRPT scheme predicts that $C_s^2$ overshoots the value $1/3$ which is the expected value at the Stefan-Boltzmann limit. The other three methods, on the other hand, predict a steady convergence toward $C_s^2=1/3$  as $T\to \infty$. Finally, Fig. \ref {Cv} which shows the specific heat as a function of $T$ for different values of $B$  
 displays a clear difference between the deconfinement (first peak) and the chiral (second peak) crossover taking place within the $T \sim 0.15-0.2 \ {\rm GeV}$ range. 
One observes a rather good agreement between the full MFIR, VMR, and SPT prescriptions even when the magnetic field reaches high values. On the other hand, the TRPT prescription predicts much lower values when compared to the other regularization schemes especially  for temperatures around and above the deconfinement pseudocritical transition. This appears to be yet another consequence of regulating  the convergent thermomagnetic integrals within this model (a byproduct of underestimating the Stefan-Boltzmann limit in the pressure). For all regularization prescriptions adopted in this work, one can observe the presence of two peaks in the specific heat as a function of the temperature: the first one (more abrupt) determines the pseudocritical temperature for deconfinement, and the second (smoother) 
determines the pseudocritical temperature for chiral symmetry restoration.  It is important to note that all of our results include IMC through $G(B)$ in the deconfinement and chiral transitions and the splitting between these pseudocritical temperatures remains almost constant if we increase the strength of the magnetic field as already reported in the context of the  SU(3) PNJL \cite{Ferreira:2014exa}. This is in the opposite behavior when compared with results of SU(3) PNJL and SU(2) LSM~\cite{Ferreira:2013oda,Mizher:2010zb}, where the splitting  increases 
with $B$. The LQCD study~\cite{Bali:2011qj} gives further support to  the  behavior  found in Ref.\cite{Ferreira:2014exa}.


\section{Conclusions}

We have considered the two-flavor PNJL model in the presence of a thermomagnetic background within the mean field framework in order to compare four different regularization prescriptions. In nonrenormalizable theories, the adoption of different regularization schemes may lead to rather different predictions when a magnetic field and thermal bath are present since one lacks further constraints such as the ones available to renormalizable theories (e.g., renormalization group equations). Apart from considering the three popular schemes represented by the SPT, TRPT, and MFIR, we have proposed an alternative procedure (dubbed VMR). Within this scheme, all divergences are first disentangled and then regulated without any further subtractions while the finite thermomagnetic contribution is integrated over the full momentum range. Comparing the behavior of different physical observables we are able to conclude that, as expected, the TRPT fails to converge to the Stefan-Boltzmann limit when high temperatures are considered. The other three prescriptions predict similar behaviors for the quark condensate, normalized pressure, speed of sound, and specific heat. Nevertheless, the MFIR displays a rather different behavior with regard to the absolute pressure and magnetization. In particular, the MFIR predictions for the latter quantity are in qualitative disagreement with the other three methods. Namely, while the SPT, TRPT, and VMR predict quark matter to be diamagnetic at the $T$, $B$ range analyzed here, the MFIR predicts it to be paramagnetic at low $B$ and diamagnetic at high field values. We believe that this different behavior is due to the missing field-dependent terms subtracted during the MFIR renormalization process. On the other hand, the results furnished by the SPT and the VMR, proposed here, are very similar both qualitatively and quantitatively. The small differences between both schemes only become apparent when examining the results for the magnetization at high field values. This probably happens because within the VMR the divergences contained in the vacuum and in the purely magnetic part have been properly isolated and regulated allowing for the sum over LL to be performed in a closed analytical form. In the view of these results, one may conclude that the VMR offers the most versatile regularization scheme to describe most observables related to magnetized quark matter.
We have also observed that, in a scenario without inverse magnetic catalysis, the paramagnetic character of QCD can be properly achieved if one makes use of the renormalized magnetization applied to the VMR scheme.

\acknowledgments

We are grateful to Konstantin Klimenko, Pedro Costa, Jo\~ao Moreira, and Norberto Scoccola for related discussions. This work was partially supported by Conselho Nacional de Desenvolvimento Cient\'ifico e Tecnol\'ogico  (CNPq), Grants No. 304758/2017-5 (R. L. S. F.), No. 304518/2019-0 (S. S. A.) and No. 303846/2017-8 (M. B. P.); Coordena\c c\~{a}o  de Aperfei\c coamento de Pessoal de  N\'{\i}vel Superior - (CAPES-Brazil)  -
Finance  Code  001 (T. E. R. and W. R. T.); Funda\c{c}\~ao de Amparo \`a Pesquisa do Estado do Rio Grande do Sul (FAPERGS), Grants No. 19/2551- 0000690-0 and No. 19/2551-0001948-3 (R. L. S. F.). The work was also part of the project Instituto Nacional de Ci\^encia e Tecnologia - F\'isica Nuclear e Aplica\c{c}\~oes (INCT -FNA) Grant No. 464898/2014-5. T. E. R. acknowledges Conselho Nacional de Desenvolvimento Cient\'{\i}fico e Tecnol\'{o}gico (CNPq-Brazil) and  Coordena\c c\~{a}o  de Aperfei\c coamento  de  Pessoal  de  N\'{\i}vel  Superior (CAPES-Brazil) for Ph.D. grants at different periods of time as well as the support and hospitality of CFisUC where part of this work was developed.

\appendix 
\newpage
\section{Vacuum Magnetic Regularization}\label{AppendixA}

Let us consider the (entangled) vacuum-magnetic parts of the thermodymanical potential as it appears in the lhs of Eq.(\ref{split1}),
\begin{equation}
 I=\sum_{f=u,d}\frac{N_c}{8 \pi^2 }\int_{\frac{1}{\Lambda^2}}^{\infty} ds \frac{e^{-M^2 s}}{s^3} \left( |q_f|B s \coth (|q_f|B s)\right),\label{term1}
\end{equation}
where we have used a simple change of variables. Note that the integral is clearly divergent for $s\to 0$. Within the VMR, we first separate the integrand of Eq.(\ref{term1}) into a divergent and a finite part \cite{Schwinger:1951nm} as $s\to 0$. With this aim, let us first expand the $\coth(|q_f|Bs)$ in Taylor series, such that

\begin{equation}\label{A2}
 \frac{e^{-M^2 s}}{s^3} \left( |q_f|B s \coth (|q_f|B s)\right)=\frac{e^{-M^2 s}}{s^3} \left(1+\frac{(|q_f|Bs)^2}{3}-\frac{(|q_f|Bs)^4}{45}+\mathcal{O}((|q_f|Bs)^6)\right), \quad (|q_f|Bs)<\pi\,.
\end{equation}
As we can see, the first two terms in Eq.(\ref{A2}) are divergent for $s\to 0$ and need regularization, so we can rewrite $I$ as 
\begin{align}
&I=I_0+I_{field}+I_{int},\label{term2}
\end{align}
where

\begin{equation}
I_0= \frac{N_fN_c}{8 \pi^2 }
  \int_{\frac{1}{\Lambda^2}}^{\infty} ds \frac{e^{-M^2 s}}{s^3}  \,
\label{ivac} 
\end{equation}
and
\begin{equation}
I_{field}= \sum_{f=u,d}N_c \frac{ (|q_f|B)^2 }{24 \pi^2 }
  \int_{\frac{1}{\Lambda^2}}^{\infty} ds \frac{e^{-M^2 s}}{s} \,.
\label{ifield}
\end{equation}
 The integral represented by Eq. (\ref{ifield}) can be further simplified by considering its explicit solution
\begin{align}
I_{field}= \sum_{f=u,d}N_c \frac{ (|q_f|B)^2 }{24 \pi^2 } \Gamma\left[0,\frac{M^2}{\Lambda^2}\right],\label{Gamma}
\end{align}
where $\Gamma\left[a,z\right]$ is the incomplete gamma function.
Expanding the above result in powers of $M^2/\Lambda^2$, one obtains
\begin{equation}
I_{field} =\sum_{f=u,d}N_c \frac{ (|q_f|B)^2 }{24 \pi^2 } \left [ -\ln\left(\frac{M^2}{\Lambda^2}\right)-\gamma_{E}+\frac{M^2}{\Lambda^2}+\mathcal{O}(M^4/\Lambda^4)
\right ] \,,
\label{Gamma_expand}
\end{equation}
which allows us to identify the divergences and finite terms as $\Lambda\to \infty$. At this limit, all mass-dependent terms vanish so that one only needs to consider
\begin{equation}
I_{field} =\sum_{f=u,d} N_c\frac{ (|q_f|B)^2 }{24 \pi^2 } \left [ -\ln\left(\frac{M^2}{\Lambda^2}\right)-\gamma_{E} \right ] \,.
\label{Gamma_expand2}
\end{equation}
This last  step guarantees that the gap equation will be consistent for the following reason. Deriving $I_{field}$ with respect to $M$, as given by 
the original Eq. (\ref {ifield}), yields  
\begin{equation}
\frac{\partial I_{field}}{\partial M} = - \sum_{f=u,d} N_c\frac{ (|q_f|B)^2 }{12 \pi^2 } M \int_{\frac{1}{\Lambda^2}}^{\infty} ds e^{-M^2 s} \,, 
\end{equation}
which would contribute to the gap equation. However, since this term is finite, one may replace $1/\Lambda^2 \to 0$ in the lower limit of the integral to get
\begin{equation}
\frac{\partial I_{field}}{\partial M} = - \sum_{f=u,d} N_c\frac{ (|q_f|B)^2 }{12 \pi^2 M } \,.
\end{equation}
This is exactly the same result that one obtains by deriving the truncated $I_{field}$ contribution, Eq. (\ref {Gamma_expand2}), with respect to $M$.  Therefore, it is Eq. (\ref  {Gamma_expand2}) and not Eq. (\ref {ifield}) the one which should be considered when evaluating the thermodynamical potential. 

The finite pure magnetic contribution $I_{int}$ is 
\begin{equation}
 I_{int}=\sum_{f=u,d}\frac{N_c}{8 \pi^2 }
  \int_0^{\infty} ds \frac{e^{-M^2 s}}{s^3} \left( |q_f|B s \coth (|q_f|B s) -1 - \frac{ (|q_f|Bs)^2 }{3}\right).\label{term3}
\end{equation}
Now, we can solve the finite integral (\ref{term3}) using the representation of the gamma function,

\begin{equation}
 \frac{\Gamma (n+1)}{(\beta)^{n+1}} =\int_0^{\infty}~ ds s^{n} e^{-\beta s}\,,
\end{equation}
and the Hurwitz-Rieman-zeta function \cite{gradshteyn2007},
\begin{align}
 \zeta\left(z,q\right)=\sum_{k=0}^{\infty}\frac{1}{\left(q+k\right)^z}\,,
\end{align}
such that

\begin{align}
 I_{int} &= \sum_{f=u,d}N_c\lim_{\epsilon \rightarrow 0} \frac{(|q_f|B)^2}{8 \pi^2 }
  \int_0^{\infty} ds~ e^{-\frac{M^2}{|q_f|B} s} s^{-3+\epsilon} \left( s \coth (s) -1 - \frac{ s^2 }{3}\right) \notag \\
&= \sum_{f=u,d}N_c\frac{(|q_f|B)^2}{8 \pi^2 } \lim_{\epsilon \rightarrow 0}
\left[ \Gamma(-1+\epsilon) \left( 2^{2-\epsilon} \zeta(-1+\epsilon,\frac{M^2}{2|q_f|B}) -
\left(\frac{M^2}{|q_f|B}\right)^{1-\epsilon}  \right) -\frac{\Gamma(-2+\epsilon)}{\left( \frac{M^2}{|q_f|B}\right)^{-2+\epsilon} }
-\frac{1}{3}\frac{\Gamma(\epsilon)}{\left( \frac{M^2}{|q_f|B}\right)^{\epsilon} }
 \right] . \label{term4}
\end{align}
Make use of some expansions such as $a^{-\epsilon} \cong 1 - \ln a ~ \epsilon + \mathcal{O}(\epsilon)$ and 

\begin{eqnarray}
 \Gamma(-n+\epsilon)=\frac{(-1)^n}{n!}\left[ \frac{1}{\epsilon} + \psi_1(n+1) + \mathcal{O}(\epsilon) \right],
\end{eqnarray}
where $\psi_1(n+1)=1+\frac{1}{2}+ ... + \frac{1}{n} - \gamma_E $, and $\gamma_E=0.577216$ is the Euler-Mascheroni constant. After some algebraic steps, we then obtain
\begin{equation}
 I_{int} = - N_c\sum_{f=u,d}\frac{(|q_f|B)^2}{2 \pi^2 }
\left[\zeta^{\prime}(-1,x_f)-\frac{1}{2}(x_f^2 -x_f)\ln x_f ~+\frac{x_f^2}{4}-\frac{1}{12}(1+\ln x_f) \right], 
\label{iqint}
\end{equation}
where we have defined $x_f=M^2/(2|q_f|B)$ so that Eq. (\ref{term2}) is exactly the rhs of Eq. (\ref{split1}).
Now adding $I_{field}$ as given by the truncated relation, Eq. (\ref{Gamma_expand2}), to  $I_{int}$ as given by Eq. (\ref {iqint}) allows us to merge two mass-dependent logarithmic terms into a mass-independent one. Namely, $-\ln (M^2/\Lambda^2) + \ln x_f = \ln [\Lambda^2/(2|q_f|B)]$ which does not contribute to the gap equation.
 Then,  Eq.(\ref{split1}) can be finally written as
\begin{equation}
\begin{split}
 \frac{N_c}{8\pi^2}\sum_{q_f=u,d}(|q_f|B)^2\int_{\frac{|q_f|B}{\Lambda^2}}^{\infty}\frac{ds}{s^2}e^{-\frac{M^2s}{|q_f|B}}\coth (s)= &\frac{N_cN_f}{8\pi^2}\int_{\frac{1}{\Lambda^2}}^{\infty}\frac{ds}{s^3}e^{-M^2s}+\frac{N_c}{24\pi^2}\sum_{q_f=u,d}(|q_f|B)^2\left[\ln\left(\frac{\Lambda^2}{2|q_f|B}\right)+1-\gamma_{E}\right]\\
 &-N_c\sum_{f=u,d}\dfrac{(|q_f|B)^2}{2\pi^2}\left[\zeta^\prime\left(-1,x_f\right)-\left[x_f^2-x_f\right]\frac{\ln x_f}{2}+\dfrac{x_f^2}{4}\right].
 \end{split}\label{int_corrected}
\end{equation}

\section{Renormalized Magnetization }\label{AppendixB}
The magnetization (${\cal M}^r$) considered in  lattice  evaluations  \cite{Bali:2013esa}, defined by Eq. (\ref {latticeREN}), is the renormalized one. However, contrary to QCD,  the PNJL model considered here represents a nonrenormalizable theory. Nevertheless, it is possible to define within the PNJL framework a quantity analogous to ${\cal M}^r$ in the following way.
First, recall that in lattice QCD applications, the thermodynamical quantities are renormalized in a way to eliminate 
the contributions of order $\mathcal{O}(eB^2)$ when $T=0$. Then, using the on shell scheme, one can define the operator 
\begin{equation}
{\cal P}[X]=\left.(eB)^2\lim_{eB\rightarrow 0}\frac{X}{(eB)^2}\right|_{T=0}\,, \label{proj} 
\end{equation}
so that, for a renormalized quantity, one has
\begin{equation}
  X^r=({1-{\cal P}})[X] \,.
\end{equation}
Within the PNJL theory, we first need to evaluate the limit $eB\rightarrow 0$ for each quantity in the VMR thermodynamical potential, Eq.(\ref{OmegaVMR}). 
In this case, the thermomagnetic contribution is obviously zero,
following the definition given by Eq. (\ref{proj}). Next, let us show that the pure magnetic contribution
also does not contribute in this limit. To this end, let us first consider the expansion of the Hurwitz zeta function, $\zeta^\prime(-1,x_f)$ \cite{Endrodi:2013cs}, in the limit $eB\rightarrow 0$. This is given by

\begin{eqnarray}\label{HurwitzB0}
 \zeta^\prime(-1,x_f)=\frac{1}{12}-\frac{x_f^2}{4}+\left (\frac{1}{12}-\frac{x_f}{2}+\frac{x_f^2}{2} \right )\ln x_f+\mathcal{O}(x_f^{-2}).
\end{eqnarray}
Substituting Eq. (\ref{HurwitzB0}) in the (finite) purely magnetic contribution $I_{int}$ given by Eq. (\ref{iqint}), one finds
\begin{eqnarray}
(eB)^2\lim_{eB\rightarrow 0} \frac{I_{int}}{(eB)^2}=0.
\end{eqnarray}
The only remaining magnetic contribution to $\Omega_{VMR}$ is given by $I_{field}$, Eq. (\ref{Gamma_expand2}).  
We then apply Eq. (\ref{proj})  to the VMR pressure, $P_{VMR}=-\Omega_{VMR}(M,\Phi,T,B)\mid_M$, obtaining our ``renormalized" pressure
\begin{eqnarray}
 P_{VMR}^r=P_{VMR}-(eB)^2\lim_{eB\rightarrow 0}\frac{P_{VMR}}{(eB)^2}\Bigr |_{T=0} \,,
\end{eqnarray}
which yields
\begin{eqnarray}
  P_{VMR}^r=P_{VMR}-\sum_{f=u,d}(|q_f|B)^2\frac{1}{8\pi^2}\left(\ln \frac{M(0)^2}{\Lambda^2}+\gamma_E\right).
\end{eqnarray}
Finally,   our ``renormalized" magnetization ${\cal M}^r$ can be directly evaluated as
\begin{equation}
 {\cal M}^r=\frac{\partial P^r_{VMR}}{\partial B} \;.
\end{equation}

\bibliographystyle{apsrev4-1}
\bibliography{PNJL_B_bib}

 \end{document}